# Electric quadrupole second harmonic generation revealing dual magnetic orders in a magnetic Weyl semimetal

Youngjun Ahn[1], Xiaoyu Guo[1], Rui Xue[2], Kejian Qu[3], Kai Sun[1], David Mandrus[2,3,4], Liuyan Zhao[1,*]

[1]Department of Physics, University of Michigan, Ann Arbor, MI, 48109, USA

[2]Department of Physics and Astronomy, University of Tennessee, Knoxville, TN, 37996, USA

[3]Department of Materials Science and Engineering, University of Tennessee, Knoxville, TN, 37996, USA

[4]Materials Science and Technology Division, Oak Ridge National Laboratory, Oak Ridge, TN, 37831, USA

[*]Email: lyzhao@umich.edu

**Broken symmetries and electronic topology are nicely manifested together in the second order nonlinear optical responses from topologically nontrivial materials. While second order nonlinear optical effects from the electric dipole (ED) contribution have been extensively explored in polar Weyl semimetals (WSMs) with broken spatial inversion (SI) symmetry, they are rarely studied in centrosymmetric magnetic WSMs with broken time reversal (TR) symmetry due to complete suppression of the ED contribution. Here, we report experimental demonstration of optical second harmonic generation (SHG) in a magnetic WSM $Co_3Sn_2S_2$ from the electric quadrupole (EQ) contribution. By tracking the temperature dependence of the rotation anisotropy (RA) of SHG, we capture two magnetic phase transitions, with both the SHG intensity increasing and its RA pattern rotating at $T_{C,1}$=175K and $T_{C,2}$=120K subsequently. The fitted critical exponents for the SHG intensity and RA orientation near $T_{C,1}$ and $T_{C,2}$ suggest that the magnetic phase at $T_{C,1}$ is a 3D Ising-type out-of-plane ferromagnetism while the other at $T_{C,2}$ is a 3D XY-type all-in-all-out in-plane antiferromagnetism. Our results show the success of detection and exploration of EQ SHG in a centrosymmetric magnetic WSM, and hence open the pathway towards the future investigation of its tie to the band topology.**

Topological Weyl semimetals (WSMs)[1,2] can only emerge when either spatial inversion (SI) or time reversal (TR) symmetry is broken and host pairs of Weyl nodes with opposite divergent Berry curvature, offering an exciting platform to explore the interplay between broken symmetries and electronic band topology. Recently, in SI-broken polar WSMs, such an interplay has been extensively investigated through the second order nonlinear optical and optoelectronic effects via the leading order electric dipole (ED) contribution[3–15] which only survives with the broken SI symmetry. In contrast, TR-broken magnetic WSMs often preserve the SI symmetry[16–18] and, therefore, fully suppress the ED contribution to the second order nonlinear effects. As a result, little effort and success have been made in exploring the nonlinear optical and optoelectronic responses in magnetic WSMs thus far, and neither has the interplay between broken symmetries and topology in this family of materials.

Very recently, it has been shown that the second order nonlinear optical effects are present in SI-preserved centrosymmetric crystals by the virtue of the next order electric quadrupole (EQ) or magnetic dipole (MD) contribution[19–25], albeit the leading order ED contribution is fully suppressed. This progress motivates us to consider EQ or MD second order nonlinear optical effects in the magnetic WSMs that break the TR symmetry while the SI symmetry is preserved. In this study, we focus on second harmonic generation (SHG) in $Co_3Sn_2S_2$ in which the frequency of the outcoming light doubles from the incoming one through the nonlinear interactions between light and electronic states inside $Co_3Sn_2S_2$ (Fig. 1a). $Co_3Sn_2S_2$ is a magnetic WSM candidate whose magnetic structure is, however, not comprehensively resolved despite the general agreement of the broken TR symmetry. Its crystal structure follows the point group $\bar{3}m$ with the in-plane crystal axis $a$ aligned along one edge of the Co kagome lattice, the other in-plane axis $b$ 120° rotated from the $a$-axis, and the out-of-plane axis $c$ coinciding with the 3-fold rotational ($C_3$) symmetry axis (Fig. 1a). Upon cooling across $T_{C,1}$ = 175 K, $Co_3Sn_2S_2$ transitions from the paramagnetic (PM) phase into a puzzling magnetic phase where the out-of-plane component of magnetic moments forms the ferromagnetic (FM) order but the in-plane component is subjected to a heated debate among i) a disordered state[26], ii) a spin glass phase[27], iii) an in-plane antiferromagnetic (AFM) phase of $\bar{3}m$ magnetic point group[28], or

iv) an AFM phase of $\bar{3}m'$ magnetic point group[29]. Upon further cooling below $T_{C,2}$ = 120 K, $Co_3Sn_2S_2$ is generally believed to develop into the fully polarized FM phase with a zero in-plane spin component[28], despite individual suggestions of ii), iii) or iv) with finite in-plane component. On top of the mystery about magnetic phases and phase transitions in $Co_3Sn_2S_2$, the domain structure and domain walls of its out-of-plane FM order further experience temperature dependent evolutions, which adds another layer of complexity in understanding the magnetism in $Co_3Sn_2S_2$. The FM domains gradually grow in lateral size over a range of temperatures between 175 K and 150 K[30,31], whereas the FM domain walls experience a phase transition at 135 K from the Néel type to the Bloch type[30]. These temperature dependent magnetic properties are summarized in Fig. 1b.

To resolve magnetic phases for both the out-of-plane and in-plane spin components, we leverage the symmetry sensitivity of SHG by performing the rotational anisotropy (RA) measurements in which the SHG intensity $I(2\omega)$ is recorded as the light-scattering-plane is rotated about the *c*-axis from the in-plane $a$-axis by an angle $\varphi$ (Fig. 1a). For the oblique incident geometry ($\theta$ = 11° in this study), there are in total four distinct polarization channels, *PP*, *PS*, *SP*, and *SS*, where *P*/*S* stands for the light polarization parallel/perpendicular to the scattering-plane and the first/second letter corresponds to the incident fundamental/reflected SHG polarization. For the normal incident configuration ($\theta$ = 0°), there are only two polarization channels, crossed and parallel, that represent the polarization relationship between incident and reflected light. The wavelength of incident fundamental light is chosen to be 800 nm, as its SHG of 400 nm matches the maximum detector sensitivity in our setup (See Methods). This choice is critical for this study because the EQ/MD SHG is typically several orders of magnitude weaker than the ED SHG if present[32].

We begin with performing RA measurements of SHG and linear responses for both normal and oblique incidence configurations to confirm the crystallographic symmetry of $Co_3Sn_2S_2$ at $T$ = 293 K (Fig. 1c). All RA SHG patterns show a $C_3$ rotational symmetry about the *c*-axis and mirror symmetry about three 120° rotated planes (marked by dotted lines in the polar plots). In consideration of the SI symmetry in the crystallographic point group $\bar{3}m$, we have simulated the functional forms of RA SHG from the EQ contribution and confirmed their excellent agreement with the experimental data, confidently showing that the EQ SHG plays the dominant role in our measurement (Supplementary Section 1). Potential SHG contributions from the surface ED, the bulk MD, and electric-field-induced SHG (EFISH) contributions show distinct RA functional forms from EQ SHG in the oblique incidence geometry channel, and hence can only contribute an insignificant weight to our data even if present (Supplementary Section 2). In addition, the RA patterns of the linear response exhibit an isotropic intensity in the parallel channels and zero signal in the crossed channels, which is consistent with the ED linear reflectance of the trigonal crystal class.

We then proceed to study the temperature dependence of SHG and linear responses, to explore their evolution across the magnetic phase transitions at $T_{C,1}$ and $T_{C,2}$. Figure 2 shows the normalized SHG intensity at $\varphi = 0^{o}$ and $\theta = 0^{o}$ with 800 nm fundamental and 400 nm SHG wavelengths and the normalized linear reflection intensity at 800 nm and 400 nm wavelengths for the parallel channel in both experiments as a function of temperature measured in a cooling cycle from 200 K to 90 K. On the one hand, the two linear response traces exhibit no detectable temperature dependencies. This observation is consistent with the previous report that there is little change in the optical conductivity at energies higher than 0.8 eV (wavelength shorter than 1550 nm) across the magnetic phase transitions[33]. On the other hand, the SHG temperature dependence clearly shows an order-parameter-like increase at $T_{C,1}$ = 175 K, and surprisingly, another anomalous upturn around $T_{C,2}$ = 120 K, which is in stark contrast to the featureless behaviors of the linear reflectivity at 400 nm and 800 nm. We further note that even in magneto-optical Kerr effect (MOKE) measurements, only the transition at $T_{C,1}$ was captured[30]. Considering that the ED linear responses at 800 nm and 400 nm are insensitive to magnetic phase transitions, we tentatively attribute the changes in the SHG temperature dependence to new SHG processes that are turned on by the emergent magnetic orders.

To reveal the nature of these new SHG processes, we compare RA SHG results at different temperatures in all polarization channels for both normal and oblique incidence geometries. Figure 3a shows representative RA SHG polar plots (top panel) and Cartesian graphs (bottom panel) at two selected temperatures, 180 K (gray circles and lines) and 90 K (blue dots and lines) that are above $T_{C,1}$ and below $T_{C,2}$, respectively. The most notable difference between SHG data at these two temperatures is a significant increase in the intensity, as clearly shown in the polar plots of all polarization channels under both incidence geometries (Fig. 3a, top panel). We highlight that, clearly distinct from the magnetic domain-sensitive magnetic circular dichroism (MCD) signal below $T_{C,2}$ (MCD mapping in Fig. 3b), the SHG intensity at 80 K is spatially homogeneous when surveying different locations of the sample and always increases from the higher temperature value (Fig. 3c). One more subtle difference is the slight rotation of the RA SHG pattern ($\Delta\varphi \approx 1^{\circ}$) between those at the two temperatures, which is better illustrated in the Cartesian graphs of normalized SHG intensity (Fig. 3a, bottom panel) for all channels as the small lateral shift between curves at the two temperatures (see examples of raw data Cartesian plots in Supplementary Section 3). This RA SHG rotation is a direct evidence of broken mirror symmetries caused by the magnetic phases. We further note that while the rotation direction is consistently the same across all the channels at one sample spot within one thermal cycle (Fig. 3a), it could alter between clockwise and counterclockwise across different sample locations and shows a positive correlation with the MCD signal sign (Fig. 3d). The contrast behavior between these two parameters, the RA SHG intensity and orientation, indicates that there are two types of new processes joining at lower temperatures, one attributed to the magnetism-induced TR invariant (*i*-type) and the other to the magnetism-induced TR broken (*c*-type)[34] (Supplementary Section 4). These two types of contributions are responsible for the intensity increase and the orientation rotation of RA SHG, respectively.

To gain insight into the two types of magnetism-induced processes across the two magnetic phase transitions, we perform systematic temperature-dependent RA SHG measurements over a temperature window ranging from 90 K to 200 K containing both $T_{C,1}$ and $T_{C,2}$. Figure 4a shows a false color map of SHG intensity as functions of polarization angle ($\varphi$) and temperature ($T$), measured in the crossed channel at the normal incidence. Throughout the whole temperature range, we note that the $C_3$ rotational symmetry is always maintained and that the SHG intensity only increases by a factor of less than 1.6, indicating that the emerging magnetic orders should preserve the $C_3$ rotational symmetry and most likely have the SI symmetry. As a result, both $i$-type and $c$-type processes belong to the EQ SHG. We can further fit RA SHG data at every individual temperature to extract the SHG intensity ($I^{2\omega}_{\text{Crossed}}(T)$) and rotation ($\Delta\varphi(T)$), whose temperature dependencies are shown in Figs. 4b and 4c, respectively. The $I^{2\omega}_{\text{Crossed}}(T)$ trace clearly captures the anomalies at two critical temperatures, $T_{C,1}$ and $T_{C,2}$. The $\Delta\varphi(T)$ trace onsets at $T_{C,1}$ and shows a further enhancement at $T_{C,2}$. Both traces support the formation of two magnetic orders, each with both $i$-type and $c$-type magnetism-induced EQ SHG contributions. Other magnetism-induced contributions (e.g., ED SHG, MD SHG, and EFISH) may be present but should be much smaller than that of EQ SHG (Supplementary Section 5).

Therefore, we establish that the two magnetic orders (with order parameters of $M_I$ and $M_{II}$) turn on both $i$-type and $c$-type EQ SHG processes, introducing $\chi^{\text{EQ}\,(i)}_{M_I}$ and $\chi^{\text{EQ}\,(c)}_{M_I}$ below $T_{C,1}$ and then further $\chi^{\text{EQ}\,(i)}_{M_{II}}$ and $\chi^{\text{EQ}\,(c)}_{M_{II}}$ below $T_{C,2}$ on top of $\chi^{\text{EQ}}_{S}$ from the crystal structure that is present at all temperatures. We can also derive the magnetic point groups for the two orders, both belonging to $\bar{3}m'$, via searching for magnetic subgroups of the high-symmetry structural point group $\bar{3}m$ that preserve both $C_3$ and SI symmetries and break the TR and mirror symmetries. We further comment that here $M_I$ and $M_{II}$ have the same magnetic point group $\bar{3}m'$ with no further symmetry reduction below $T_{C,2}$, under the analysis based on the crystallographic point group $\bar{3}m$. If we were to analyze based on the kagome lattice of Co magnetic sites, we will see the subsequent symmetry reductions below both $T_{C,1}$ and $T_{C,2}$ (Supplementary Section 6). We note that subgroups of $\bar{3}m'$ with lower symmetries would, in principle, be consistent with our results. We can then simulate the functional forms for RA SHG at the three temperature ranges: above $T_{C,1}$, with only structural contribution ($\chi^{\text{EQ}}_{S}$); between $T_{C,1}$ and $T_{C,2}$, with the coherent interference between structural and $M_I$-induced contributions ($\chi^{\text{EQ}}_{S}$; $\chi^{\text{EQ}\,(i)}_{M_I}$ and $\chi^{\text{EQ}\,(c)}_{M_I}$); and below $T_{C,2}$, with the coherent superposition of structural, $M_I$-, and $M_{II}$-induced contributions ($\chi^{\text{EQ}}_{S}$; $\chi^{\text{EQ}\,(i)}_{M_I}$ and $\chi^{\text{EQ}\,(c)}_{M_I}$; $\chi^{\text{EQ}\,(i)}_{M_{II}}$ and $\chi^{\text{EQ}\,(c)}_{M_{II}}$) (Supplementary Section 4). The experimental data in Fig. 3a and Fig. 4a are well fitted by the simulations. We note that the $I^{2\omega}_{\text{Crossed}}$ increases to ~ 1.5 times whereas $\Delta\varphi$ only changes by ~1°, showing that $\chi^{\text{EQ}}_{\text{S}} > \chi^{\text{EQ}\,(i)}_{\text{M}_{\text{I,II}}} \gg \chi^{\text{EQ}\,(c)}_{\text{M}_{\text{I,II}}}$.

To understand the $I^{2\omega}_{\text{Crossed}}(T)$ and $\Delta\varphi(T)$ traces in Figs. 4b and 4c, we take the leading order approximation based on the knowledge of $\chi^{\text{EQ}}_{\text{S}} > \chi^{\text{EQ}\,(i)}_{\text{M}_{\text{I,II}}} \gg \chi^{\text{EQ}\,(c)}_{\text{M}_{\text{I,II}}}$ (Supplementary Section 4). For $I^{2\omega}_{\text{Crossed}}(T)$, the SHG intensity above $T_{\text{C},1}$ is from $\chi^{\text{EQ}}_{\text{S}}$ that is temperature independent. The increase in the SHG intensity below $T_{\text{C},1}$ mainly results from $\chi^{\text{EQ}\,(i)}_{\text{M}_{\text{I}}}$ that scales quadratically with $\text{M}_{\text{I}}$ while the additional enhancement below $T_{\text{C},2}$ is primarily contributed by $\chi^{\text{EQ}\,(i)}_{\text{M}_{\text{II}}}$ that is proportional to $\text{M}_{\text{II}}^2$. As $\text{M}_{\text{I}}$ and $\text{M}_{\text{II}}$ are order parameters of second order phase transitions, their temperature dependencies are described[19] by $\text{M}_{\text{I}} \propto |T\text{-}T_{\text{C},1}|^{\beta_{\text{C},1}}$ and $\text{M}_{\text{II}} \propto |T\text{-}T_{\text{C},2}|^{\beta_{\text{C},2}}$, leading to the functional form of $I^{2\omega}_{\text{Crossed}}(T) = A + B|T\text{-}T_{\text{C},1}|^{2\beta_{\text{C},1}} + C|T\text{-}T_{\text{C},2}|^{2\beta_{\text{C},2}}$, with $\beta_{\text{C},1}$ and $\beta_{\text{C},2}$ being critical exponents for $\text{M}_{\text{I}}$ and $\text{M}_{\text{II}}$, respectively, and A, B, and C are weights of the crystallographic, $\text{M}_{\text{I}}$-, and $\text{M}_{\text{II}}$-induced contributions to EQ SHG. Fitting this functional form of $I^{2\omega}_{\text{Crossed}}(T)$ to the data in Fig. 4b, we obtain $2\beta_{\text{C},1} = 0.63\pm0.03$ that agrees with a critical exponent $\beta = 0.32$ across $T_{\text{C},1}$ from MOKE experiments[30] and $2\beta_{\text{C},2} = 0.71\pm0.04$ that is close to $\beta = 0.35$ for 3D XY-type order parameter[35]. A similar but independent fit to $\Delta\varphi(T)$ in Fig. 4c, primarily contributed by $\chi^{\text{EQ}\,(c)}_{\text{M}_{\text{I}}}$ and $\chi^{\text{EQ}\,(c)}_{\text{M}_{\text{II}}}$ scaling linearly with $\text{M}_{\text{I}}$ and $\text{M}_{\text{II}}$, respectively, results in $\beta_{\text{C},1} = 0.31\pm0.08$ and $\beta_{\text{C},2} = 0.35\pm0.11$ that are consistent with the extracted values from $I^{2\omega}_{\text{Crossed}}(T)$.

Gathering together what we have learnt about the magnetic point groups, the *i*- and *c*-type EQ SHG, and the critical exponents, we can then comment on the nature of the two magnetic orders. The first magnetic order emergent at $T_{\text{C},1}$ based on our RA SHG analysis is largely consistent with the literature[16,18,28,36], which is an easy-axis (Ising-type) FM order formed by the out-of-plane spin component. The second magnetic order onset at $T_{\text{C},2}$ captured by our RA SHG without an external magnetic field is unexpected, because it is known to be rather subtle or hidden in resistivity[27,36], anomalous Hall[16], MOKE[30], and muon spin rotation spectroscopy[28] measurements. This observation of a further upturn anomaly of the RA SHG rotation at $T_{C,2}$ convinces us the broken mirror symmetry for this second magnetic phase and therefore resolves its magnetic point group to be $\bar{3}m'$ (or lower), whereas the fitted critical exponent around 0.35 suggests its XY-type nature. These results rule out the debated candidates including spin glass where the TR and mirror symmetries are expected[27], domain walls for which the $C_3$ and SI symmetries should break [30], in-plane AFM of $\bar{3}m$ that has a distinct magnetic point group[28], and fully polarized out-of-plane FM that should be of 3D Ising-type. Consequently, the most plausible possibility is the in-plane AFM of $\bar{3}m'$ as depicted in the option iv)[29] of Fig. 1b. Taking account of all results, the development of the magnetism summarized from our RA SHG is shown in Fig. 4d. The paramagnetic phase above $T_{C,1}$ is transformed into the FM order with the out-of-plane spin component across $T_{C,1}$ while the in-plane spin component coexists but fluctuates strongly[28]. Upon further cooling across $T_{C,2}$, the in-plane spin component become static and develop the AFM order of $\bar{3}m'$. A 3D cartoon illustration of the magnetic phase in shown in Fig. 4e, where the spins are primarily aligned ferromagnetically along the *c*-axis but are canted a bit to arrange antiferromagnetically within the *a*-*b* plane.

To conclude, we have successfully detected the EQ SHG from a magnetic WSM $Co_3Sn_2S_2$, further leveraged its polarization and temperature dependencies to capture two magnetic phase transitions, and eventually managed to pin down the nature of magnetism across both transitions. It successfully demonstrates the advancement of using nonlinear optics to study nontrivial magnetic textures, even those with centrosymmetry, representing a further step forward from detecting complex electric dipolar textures and paving the way towards reliably investigating multipolar orders. Looking forward, we foresee multiple additional opportunities. First, it would be insightful to explore the relationship between the out-of-plane FM and in-plane AFM for providing a comprehensive picture of magnetic domains and domain walls in this kagome magnetic system. Second, it could be fruitful to examine the evolution of electronic band topology across $T_{C,1}$ and $T_{C,2}$, to shed light on the interplay between magnetism and topology. Third, the photon energy involved in the current study is beyond the Weyl cone of $Co_3Sn_2S_2$. Therefore, the EQ SHG processes perhaps mostly concern about topologically trivial bands, and our focus here is about broken symmetries rather than band topology. Yet, it remains as an open theoretical question whether the magnitude of $\chi^{\mathrm{EQ}}$ can be expressed in relation to the Berry curvature at Weyl nodes, and also a challenging experimental adventure of how EQ SHG or EQ photocurrent would turn out with lower photon energies ranging from near the Weyl points to within the Weyl cones.

## Methods

### RA SHG and linear experiments

The RA optical response measurements were conducted under a geometry shown in Fig. 1a with a femtosecond light source with a frequency of 200 kHz and a pulse duration of 40 fs. The wavelength was tuned to 800 nm using a tunable optical parametric amplifier. A BBO crystal was used to generate 400 nm-fundamental light from 800 nm-light through a frequency doubling process. The incident light was focused on a sample surface to a 30 μm diameter with a fluence of 0.5 mJ $cm^{-2}$ and 0.1 mJ $cm^{-2}$ for SHG and linear reflectance measurements, respectively. The RA measurements were conducted based on a rotating scattering-plane technique[37]. Reflected fundamental (800 nm and 400 nm) were collected by a photodiode. The photodiode for the linear reflection experiments was connected to a lock-in amplifier, and the demodulated signal with 200 kHz was measured. The intensity of the SHG light (400 nm) was measured by EMCCD camera in front of which a set of edge-pass filters are used to selectively allow the 400 nm-light to the camera while the transmission of the light with different wavelengths is suppressed. All experiments were performed with a pressure $< 5\times10^{-7}$ mbar.

### Sample Growth

$Co_3Sn_2S_2$ single crystals were grown from excess tin using the self-flux method[38]. Cobalt slug (Alfa Aesar, 99.995%), sulfur pieces (Alfa Aesar, 99.9995%), and tin shot (Alfa Aesar, 99.99+%) with an atomic ratio of Co:S:Sn = 9:8:83 were placed in a 2 ml $Al_2O_3$ Canfield crucible set[39] and then sealed under vacuum in a silica tube. The tube was heated to 400°C at 100 °C/hour. After dwelling for 4 hours, the tube was heated to 1100 °C at the same rate and kept at this temperature for 24 hours. After that, the tube was cooled to 700 °C at 3 °C/hour. Finally, the tube was inverted and centrifuged in order to separate the crystals from the flux.

## Acknowledgements

We acknowledge the valuable discussions with Igor Mazin and Lu Li. L.Z. acknowledges support by AFOSR YIP grant no. FA9550-21-1-0065, NSF CAREER grant no. DMR-174774, and Alfred P. Sloan Foundation. K.S. acknowledges the support by the Office of Navy Research grant no. N00014-21-1-2770 and the Gordon and Betty Moore Foundation grant no. N031710. D.M. acknowledges the support of AFSOR MURI grant FA9550-20-1-0322.

## Contributions

Y.A. and L.Z. conceived the project. Y.A. planned and performed the experiment under the advice by L.Z. Y.A. and X.G. performed the point group symmetry simulation under the guidance of L.Z. R.X., K.Q. and M.D. grew and characterized the samples. Y.A., K.S., and L.Z. analyzed the data. Y.A. and L.Z. wrote and revised the manuscript with comments from all authors.

## Figures

### Figure 1

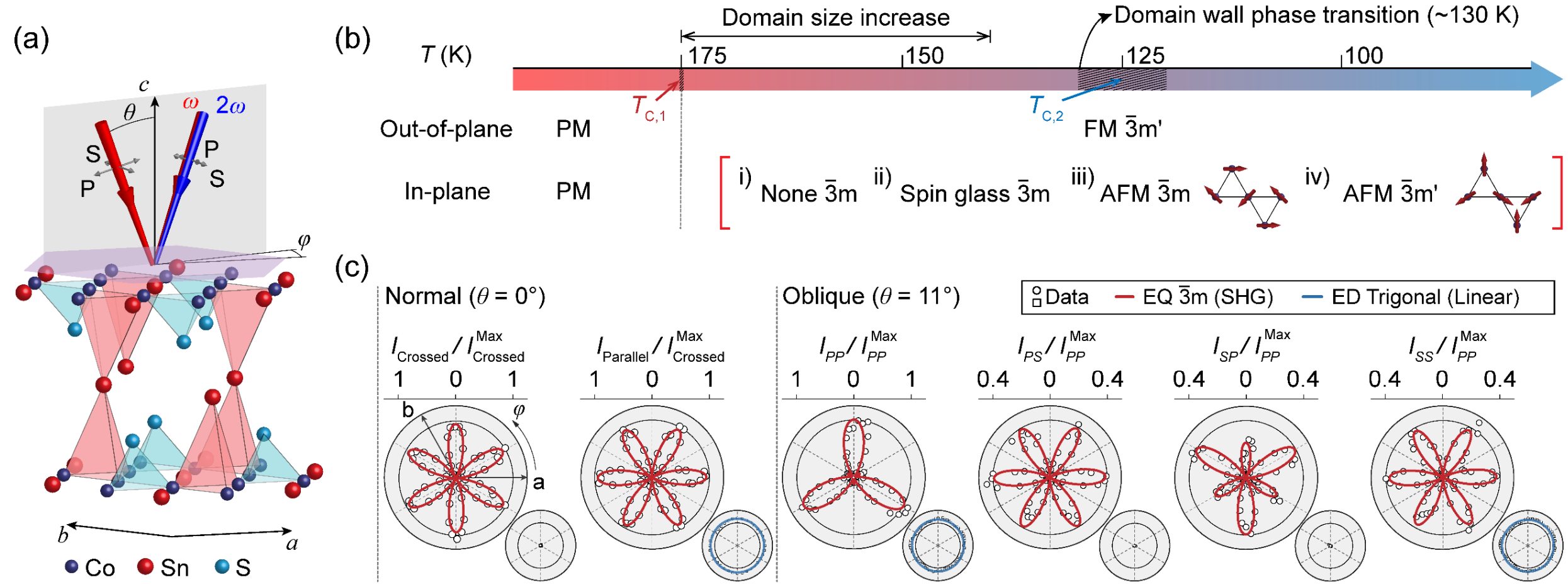

**Fig. 1** (a) Illustration of optical rotational anisotropy experiment on a magnetic WSM $Co_3Sn_2S_2$. The polarization of incident and reflected beams at the normal and oblique incidence ($\theta = 0°$ and $11°$) is selected to be parallel ($P$) or perpendicular ($S$) to the scattering plane (grey-shaded plane). The reflected beam intensity at the fundamental and SHG frequencies ($\omega$ and $2\omega$) is measured as a function of the angle $\varphi$ between the crystallographic $a$–$c$ plane and the scattering plane. (b) Summary of magnetic phases and phase transitions in $Co_3Sn_2S_2$ across two critical temperatures $T_{C,1} = 175$ K and $T_{C,2} \approx 125$ K with proposed point-group notations. (c) Rotation anisotropy of EQ SHG response and ED linear response obtained at normal and oblique incidence and $T = 293$ K. The data are fitted to the simulated EQ SHG response (red) under the $\bar{3}m$ point group and ED linear response (blue) of the trigonal crystal system. Crossed and parallel channels for the normal incidence are measured with the polarization of incident beam perpendicular and parallel to that of reflected beam. *PP*, *PS*, *SP*, and *SS* denote the selected polarizations of the incident (former) and reflected (latter) light in oblique incident experiments.

**Figure 2**

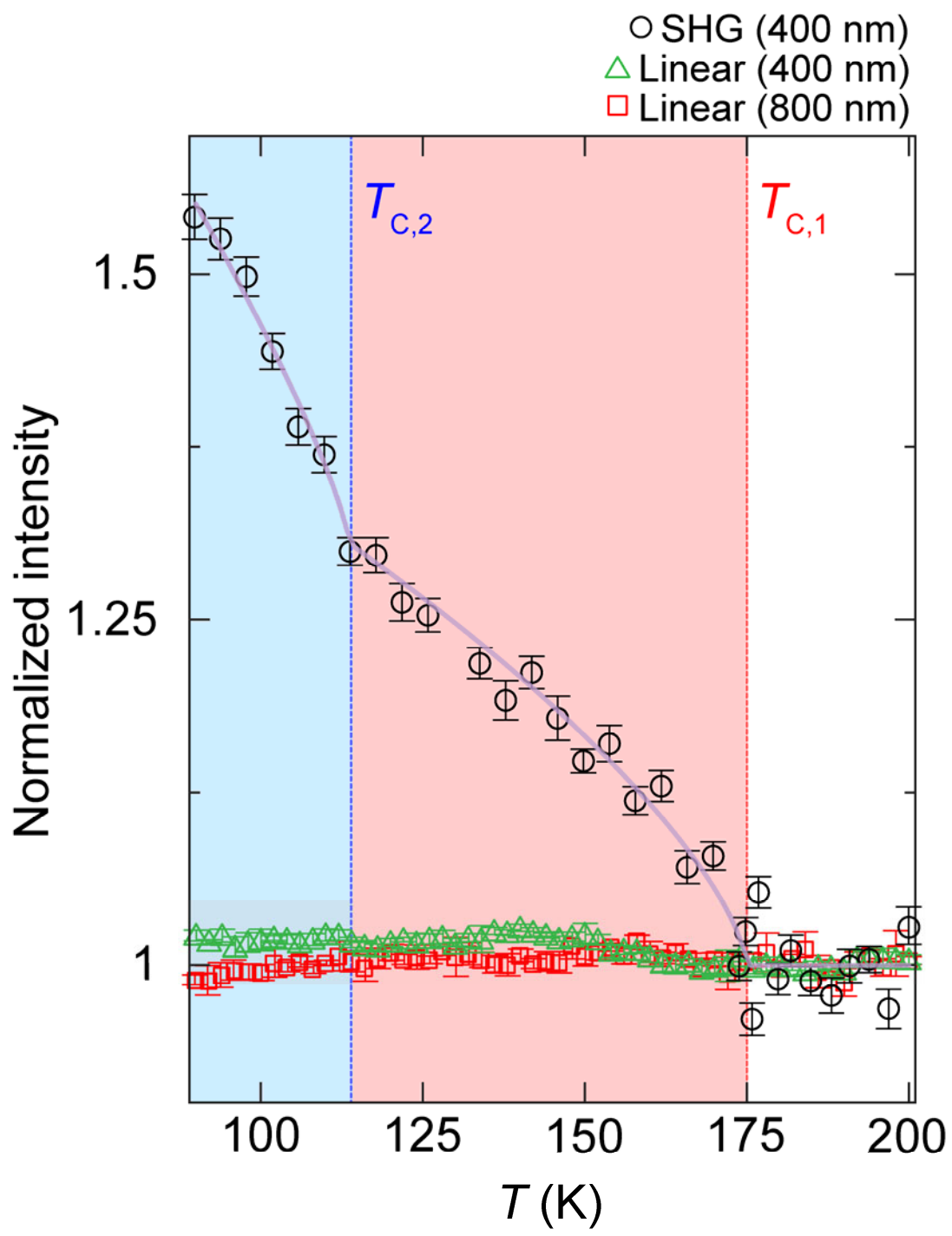

**Fig. 2** Zero-field temperature-dependent measurements of the SHG response with a fundamental (SHG) wavelength of 800 nm (400 nm) and the linear response of 400 nm and 800 nm wavelengths. All data were obtained at the normal incidence geometry ($\theta = 0°$) in the parallel polarization channel with polarizations aligned along the *a*-axis ($\varphi = 0°$). The measured intensities are normalized to the values measured at $T = 293$ K. The transition temperature $T_{C,1}$ is marked with the red dashed line. The blue dashed line indicates the secondary transition temperature $T_{C,2}$ near 120 K deduced from the SHG data.

Figure 3

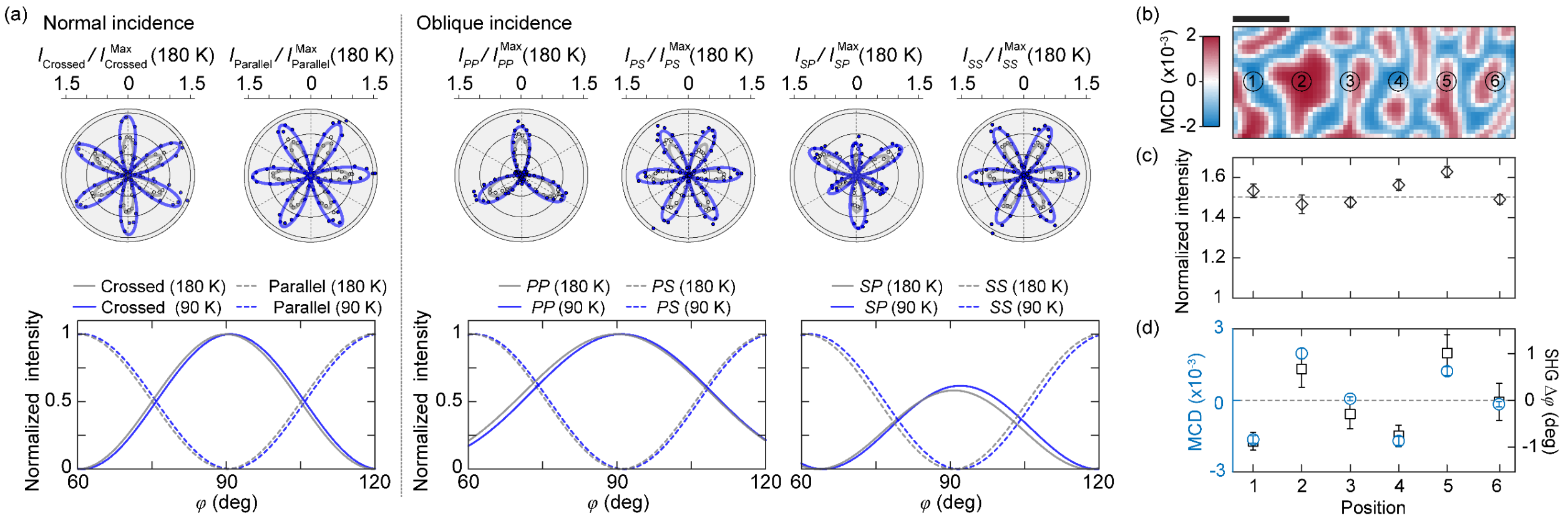

**Fig. 3** (a) (Top panel) Polar plots of RA SHG data measured in all polarization channels at 180 K (grey lines) and 90 K (blue lines) for the normal and oblique incidence geometries. The intensities are normalized by the maximum intensity of each channel at 180 K. Bottom panel displays fits to RA SHG data as a function of $\varphi$ using the derived functional forms of EQ SHG response under the $\bar{3}m'$ magnetic point group. The fitted lines are normalized by their own maximum values to show the angular rotation of the RA SHG patterns below the transition temperatures. (b) A scanning map of MCD measured at $T = 80$ K with a scale bar of 50 μm. The numbered locations are where the SHG RA patterns in the crossed channel are measured above $T_{C,1}$ ($T = 200$ K) and below $T_{C,2}$ ($T = 80$ K). (c) The intensities of the SHG RA patterns at the numbered locations below $T_{C,2}$ ($T = 80$ K), normalized by the intensities at each location above $T_{C,1}$ ($T = 200$ K). (d) The rotation of the SHG RA patterns $\Delta\varphi$ at $T = 80$ K with respect to the RA patterns at $T = 200$ K. The signs of $\Delta\varphi$ show a positive correlation with those of MCD values measured at $T = 80$ K.

**Figure 4**

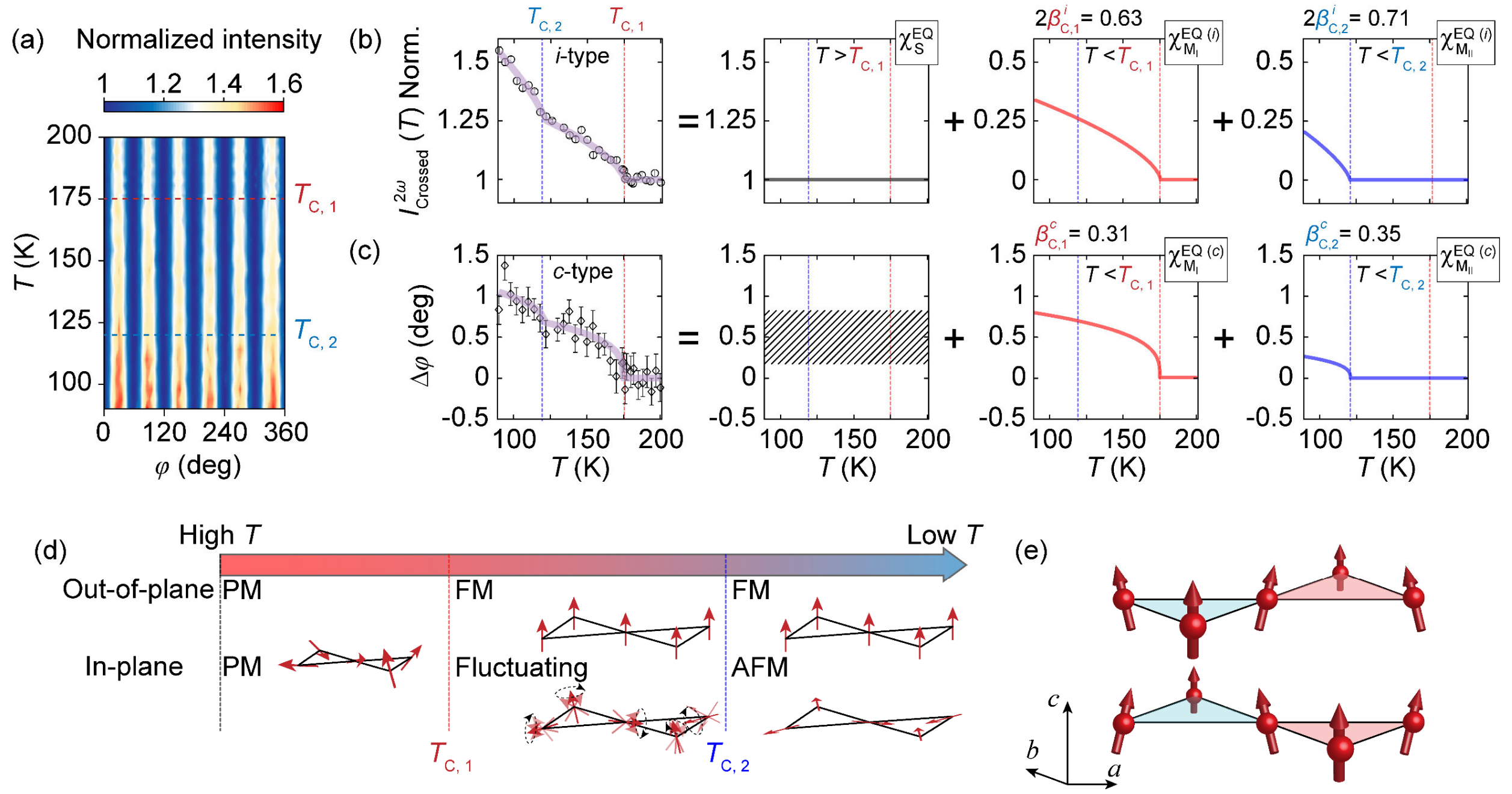

**Fig. 4** (a) Angular- and temperature-dependent RA SHG intensity obtained in the crossed polarization channel at normal incidence. Temperature dependence of the SHG intensity (b) and angular rotation $\Delta\varphi$ (c) from the TR invariant *i*-type and the TR broken *c*-type contribution, respectively. $T_{C,1}$ and $T_{C,2}$ are marked by dashed red line and blue line, respectively. Above $T_{C,1}$, there is temperature-independent intensity from the crystallographic contribution $\chi_S^{\mathrm{EQ}}$. Upon the transition into the first magnetic order ($M_I$) below $T_{C,1}$, TR invariant *i*-type $\chi_{\mathrm{M_I}}^{\mathrm{EQ}\,(i)}$ and TR broken *c*-type $\chi_{\mathrm{M_I}}^{\mathrm{EQ}\,(c)}$contributions lead to the change in the SHG intensity and $\Delta\varphi$, respectively, indicating a phase transition with a critical exponent $\beta_{C,1} \approx 0.31$. On further cooling below $T_{C,2}$, the SHG intensity and $\Delta\varphi$ shows the additional upturn with a different exponent value of $\beta_{C,2} \approx 0.35$, resulting from another set of contributions, $\chi_{\mathrm{M_{II}}}^{\mathrm{EQ}\,(i)}$and $\chi_{\mathrm{M_{II}}}^{\mathrm{EQ}\,(c)}$, of a second magnetic order ($M_{II}$). (d) A diagram of magnetic phases across $T_{C,1}$ and $T_{C,2}$ with spin components along the out-of-plane and in-plane directions. (e) The spin configuration below $T_{C,2}$ illustrating the dominant FM spins along the *c*-axis with the in-plane AFM canted moment within the *a*-*b* plane.

Supplementary Information for

# Electric quadrupole second harmonic generation revealing dual magnetic orders in a magnetic Weyl semimetal

Youngjun Ahn[1], Xiaoyu Guo[1], Rui Xue[2], Kejian Qu[3], Kai Sun[1], David Mandrus[2,3,4], Liuyan Zhao[1,*]

[1]Department of Physics, University of Michigan, Ann Arbor, MI, 48109, USA

[2]Department of Physics and Astronomy, University of Tennessee, Knoxville, TN, 37996, USA

[3]Department of Materials Science and Engineering, University of Tennessee, Knoxville, TN, 37996, USA

[4]Materials Science and Technology Division, Oak Ridge National Laboratory, Oak Ridge, TN, 37831, USA

[*]Corresponding author: lyzhao@umich.edu

## Table of Contents

**Section 1. Functional forms of EQ SHG of crystallographic $\bar{3}m$ and $i$-type $\bar{3}m'$**

First, in order to obtain non-zero tensor elements for EQ SHG under $\bar{3}m$ point group under normal and oblique incidence geometries, we derived a matrix for a polar tensor[1] using an equation:

$$\chi_{ijkl} = \sigma_{ip}\sigma_{jq}\sigma_{kr}\sigma_{ls}\chi_{pqrs}$$

where $\chi$ is the polar EQ susceptibility tensor and $\sigma$ is a generating matrix for each point group. This equation can be applicable for the derivation of tensor elements for both structural $\bar{3}m$ and $i$-type time-reversal invariant $\bar{3}m'$. The number of derived independent non-zero tensor elements[2] is 14. An effective nonlinear polarization $P$ is expressed by[3]

$$P_i(2\omega) \propto \chi_{ijkl}E_j(\omega)\nabla_k E_l(\omega)$$

where $E$ is the electric field of incoming light. The additional constraint of degenerate SHG symmetries ($j = l$) reduces the non-zero tensor elements to 11 as shown below[4]

$$\chi_{ijkl}^{EQ(s\text{ and }i)} = \begin{pmatrix} \begin{pmatrix} xxxx & 0 & 0 \\ 0 & xxyy & -yyyz \\ 0 & -yyzy & xxzz \end{pmatrix} & \begin{pmatrix} 0 & xyxy & -yyyz \\ xxyy & 0 & 0 \\ -yyzy & 0 & 0 \end{pmatrix} & \begin{pmatrix} 0 & -yyyz & xzxz \\ -yyyz & 0 & 0 \\ xxzz & 0 & 0 \end{pmatrix} \\ \begin{pmatrix} 0 & xxyy & -yyyz \\ xyxy & 0 & 0 \\ -yyzy & 0 & 0 \end{pmatrix} & \begin{pmatrix} xxyy & 0 & 0 \\ 0 & xxxx & yyyz \\ 0 & yyzy & xxzz \end{pmatrix} & \begin{pmatrix} -yyyz & 0 & 0 \\ 0 & yyyz & xzxz \\ 0 & xxzz & 0 \end{pmatrix} \\ \begin{pmatrix} 0 & -zyyy & zzxx \\ -zzyy & 0 & 0 \\ zxzx & 0 & 0 \end{pmatrix} & \begin{pmatrix} -zyyy & 0 & 0 \\ 0 & zyyy & zzxx \\ 0 & zxzx & 0 \end{pmatrix} & \begin{pmatrix} zzxx & 0 & 0 \\ 0 & zzxx & 0 \\ 0 & 0 & zzzz \end{pmatrix} \end{pmatrix}$$

Then, we applied a transformation of the tensor elements into the rotation frame of the scattering plane using the rotation matrix $R$ about the crystallographic $c$-axis to simulate the rotational anisotropy of the second harmonic generation using a following equation[3]

$$\chi_{i'j'k'l'}^{EQ}(\varphi) = R_{i'i}R_{j'j}R_{k'k}R_{l'l}\chi_{ijkl}^{EQ}$$

Finally, the functional forms for the EQ SHG intensity are derived from the equation

$$I^{2\omega}(\varphi) \propto \left|\hat{e}_{i'}(2\omega)\chi_{i'j'k'l'}^{EQ}(\varphi)\hat{e}_{j'}(\omega)\partial_{k'}\hat{e}_{l'}(\omega)\right|^2 I(\omega)^2$$

where $\hat{e}$ is the polarization of the incident and reflected light, and $I(\omega)$ is the intensity of the incoming light. By selecting the polarization of the incident and reflected beam to be crossed or parallel for the normal incidence geometry and $P$ or $S$ for the oblique incidence geometry, the functional forms under a series of polarization geometries are given below.

$$I_{PP}^{2\omega}(\varphi) \propto \cos^2(\theta)\left\{\sin^2(\theta)\left[\chi_{zxzx}\cos^2(\theta) + (2\chi_{zzxx} + \chi_{zzzz})\sin^2(\theta) + \chi_{zyyy}\cos(\theta)\sin(\theta)\sin(3\varphi)\right]^2 + \left[\left(2\chi_{xxyy} + 2\chi_{xxzz} + \chi_{xyxy}\right)\cos^2(\theta)\sin(\theta) + \chi_{xzxz}\sin^3(\theta) + \chi_{yyzy}\cos^3(\theta)\sin(3\varphi) + \chi_{yyzy}\sin(\theta)\sin(2\theta)\sin(3\varphi)\right]^2\right\}$$

$$I_{PS}^{2\omega}(\varphi) \propto \cos^2(\theta)\cos^2(3\varphi)\left[\chi_{yyzy}\cos^2(\theta) + 2\chi_{yyyz}\sin^2(\theta)\right]^2$$

$$I_{SP}^{2\omega}(\varphi) \propto \cos^2(\theta)\left[\chi_{xyxy}\sin(\theta) - \chi_{yyzy}\cos(\theta)\sin(3\varphi)\right]^2 + \sin^2(\theta)\left[\chi_{zxzx}\cos(\theta) - \chi_{zyyy}\sin(\theta)\sin(3\varphi)\right]^2$$

$$I_{SS}^{2\omega}(\varphi) \propto \chi_{yyzy}^2\cos^2(\theta)\cos^2(3\varphi)$$

$$I_{\mathrm{Crossed}}^{2\omega}(\varphi) \propto \chi_{yyzy}^2\sin^2(3\varphi)$$

$$I_{\mathrm{Parallel}}^{2\omega}(\varphi) \propto \chi_{yyzy}^2\cos^2(3\varphi)$$

where $\theta = 11°$ for the experiments performed in this work. All simulated SHG RA patterns are displayed in Fig. S1.

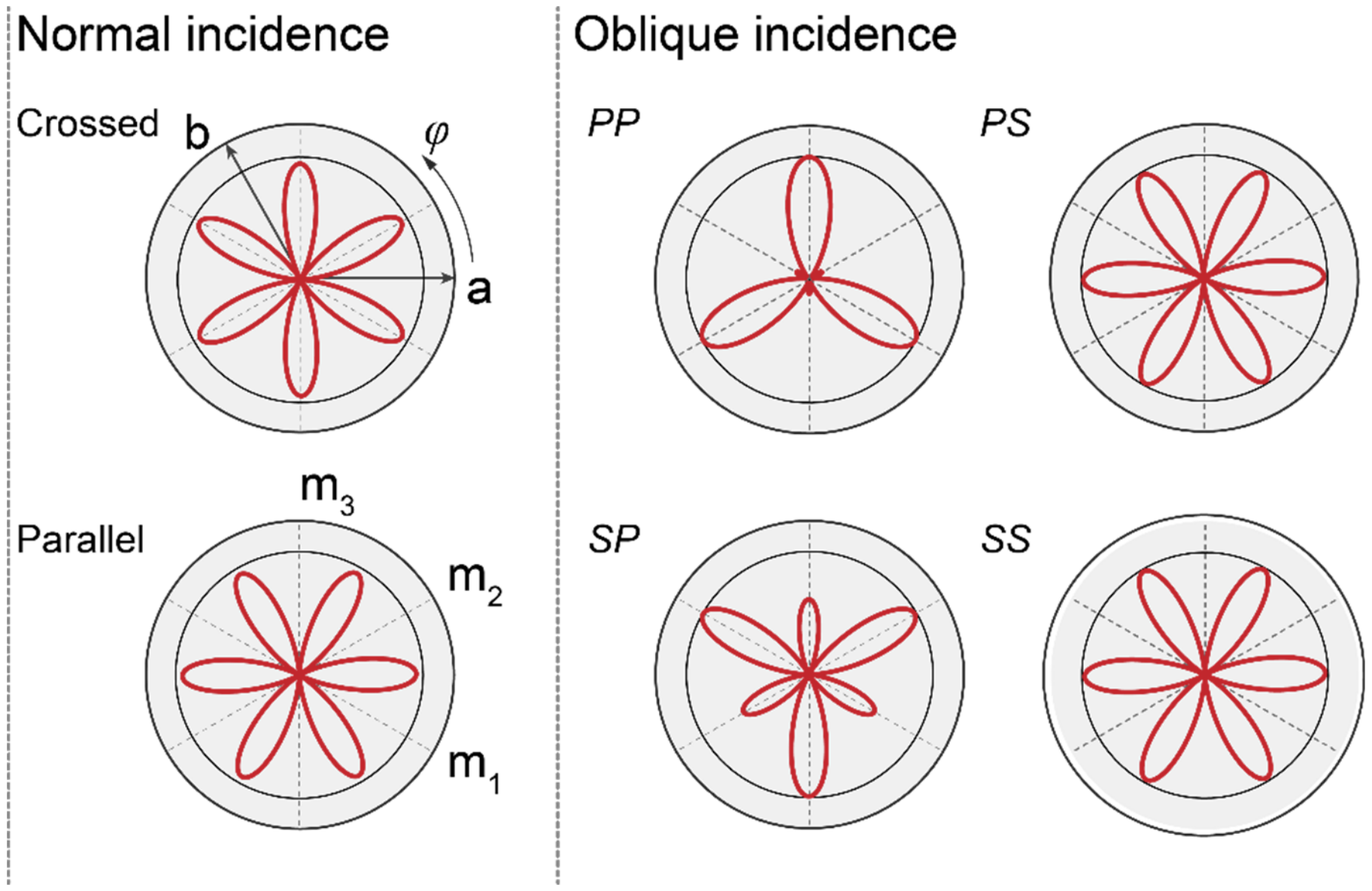

**Figure S1.** Simulated EQ SHG for crystallographic $\bar{3}m$ and *i*-type $\bar{3}m'$ for the normal incidence (left) and the oblique incidence (right) geometries. The rotation of the scattering plane is simulated by changing an angle $\varphi$ about the *c*-axis. The three mirror planes $m_1$, $m_2$, and $m_3$ are indicated by gray dashed lines. The alternating RA SHG lobes are observed in *PP* and *SP* channels while all other channels show six even lobes.

**Section 2. Functional forms of surface ED and bulk MD SHG under the $\bar{3}m$ and $\bar{3}m'$**

**Surface ED SHG of $3m$ and $3m'$ (*i*-type)**

While the ED contribution to SHG is forbidden due to the preserved inversion symmetry in the $\bar{3}m$, the surface ED contribution is plausible due to an absence of the inversion symmetry at the surface, resulting in the reduced point group to $3m$. Following the same procedure in section 1, we obtained the non-zero tensor elements for time-invariant *i*-type ED contribution as below. Here, the degenerate SHG symmetries (*j* = *k*) are considered to reduce the number of the non-zero tensor elements.

$$\chi_{ijk}^{ED} = \begin{pmatrix} \begin{pmatrix} 0 \\ -xyx \\ yyz \end{pmatrix} & \begin{pmatrix} -xyx \\ 0 \\ 0 \end{pmatrix} & \begin{pmatrix} yyz \\ 0 \\ 0 \end{pmatrix} \\ \begin{pmatrix} -xyx \\ 0 \\ 0 \end{pmatrix} & \begin{pmatrix} 0 \\ xyx \\ yyz \end{pmatrix} & \begin{pmatrix} 0 \\ yyz \\ 0 \end{pmatrix} \\ \begin{pmatrix} zyy \\ 0 \\ 0 \end{pmatrix} & \begin{pmatrix} 0 \\ zyy \\ 0 \end{pmatrix} & \begin{pmatrix} 0 \\ 0 \\ zzz \end{pmatrix} \end{pmatrix}$$

The transformed matrix $\chi_{i'j'k'}^{ED}$ into the rotated frame is used to derive the intensity of SHG RA as a function of $\varphi$, following an equation

$$I^{2\omega}(\varphi) \propto \left| \hat{e}_{i'}(2\omega) \chi_{i'j'k'}^{ED}(\varphi) \hat{e}_{j'}(\omega) \hat{e}_{k'}(\omega) \right|^2 I(\omega)^2$$

With the equation for SHG intensities, the functional forms for all polarization channels are given below.

$$I_{PP}^{2\omega}(\varphi) \propto \left[\chi_{zyy}^2 \sin^2(\theta) + \chi_{zzz}^2 \sin^3(\theta)\right]^2 + \cos^4(\theta) \left(2\chi_{yyz} \sin(\theta) + \chi_{xyx} \cos(\theta) \sin(3\varphi)\right]^2$$

$$I_{PS}^{2\omega}(\varphi) \propto \chi_{xyx}^2 \cos^4(\theta) \cos^2(3\varphi)$$

$$I_{SP}^{2\omega}(\varphi) \propto \chi_{zyy}^2 \sin^2(\theta) + \chi_{xyx}^2 \cos^2(\theta) \sin^2(3\varphi)$$

$$I_{SS}^{2\omega}(\varphi) \propto \chi_{yxy}^2 \cos^2(3\varphi)$$

$$I_{\text{Crossed}}^{2\omega}(\varphi) \propto \chi_{yxy}^2 \sin^2(3\varphi)$$

$$I_{\text{Parallel}}^{2\omega}(\varphi) \propto \chi_{yxy}^2 \cos^2(3\varphi)$$

In the *SP* channel, the intensity of SHG RA pattern is predicted to exhibit six even lobes as a function of $\varphi$, which is not consistent with the alternating lobes in the *SP* channel for EQ $\bar{3}m$ as shown in Fig. S2. This result allows us to rule out the possibility of surface ED contributions.

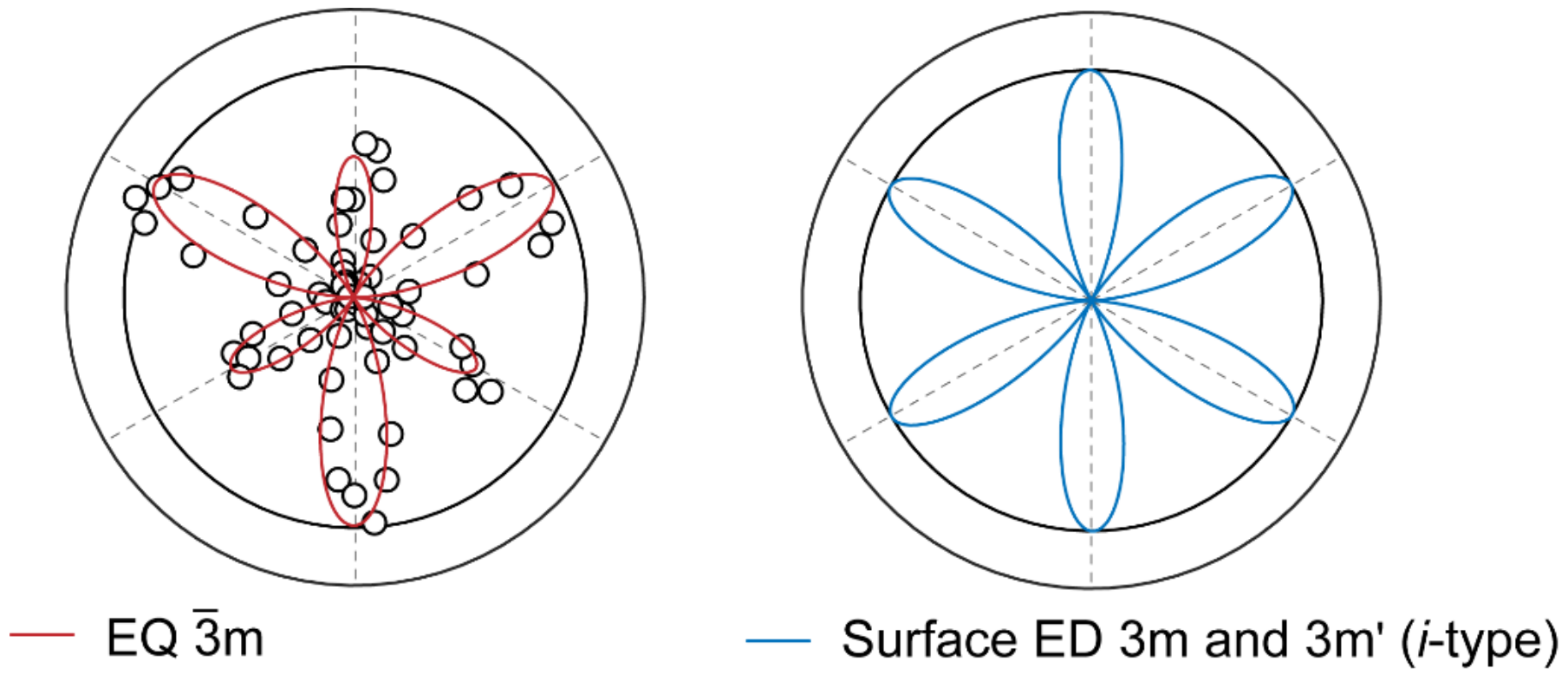

**Figure S2.** SHG RA data for the *SP* channel fitted with the simulated *SP* channel of EQ SHG (red) and SHG RA pattern simulated for time-invariant *i*-type surface ED contribution (blue)

**Surface ED SHG of $3m'$ (*c*-type):**

For the time-reversal noninvariant (broken) *c*-type surface ED contribution, the polar tensor elements for nonlinear susceptibility can be derived using[1]

$$\chi_{ijk} = (-1)\sigma_{ip}\sigma_{jq}\sigma_{kr}\chi_{pqr}$$

Here, this equation is only applicable for a generating matrix $\sigma$ for operations including the time-reversal operator $m'$ in the $\bar{3}m'$ point group. Other generating matrixes should be applied following the equation for the *i*-type surface ED SHG. Considering the degenerate symmetries ($j = k$),the obtained susceptibility tensor elements are

$$\chi_{ijk}^{ED} = \begin{pmatrix} \begin{pmatrix} yxy \\ 0 \\ 0 \end{pmatrix} & \begin{pmatrix} 0 \\ -yxy \\ yxz \end{pmatrix} & \begin{pmatrix} 0 \\ yxz \\ 0 \end{pmatrix} \\ \begin{pmatrix} 0 \\ -yxy \\ -yxz \end{pmatrix} & \begin{pmatrix} -yxy \\ 0 \\ 0 \end{pmatrix} & \begin{pmatrix} -yxz \\ 0 \\ 0 \end{pmatrix} \\ \begin{pmatrix} 0 \\ zyx \\ 0 \end{pmatrix} & \begin{pmatrix} -zyx \\ 0 \\ 0 \end{pmatrix} & \begin{pmatrix} 0 \\ 0 \\ 0 \end{pmatrix} \end{pmatrix}$$

Employing the procedures to transform the matrix into the rotated frame, we obtained the functional forms for all polarization channels used in the experiment as described below.

$$I_{PP}^{2\omega}(\varphi) \propto \chi_{yxy}^2 \cos^6(\theta)\cos^2(3\varphi)$$

$$I_{PS}^{2\omega}(\varphi) \propto \left[-2\chi_{yzx}\sin(\theta)\cos(\theta) + \chi_{yxy}\cos^2(\theta)\sin(3\varphi)\right]^2$$

$$I_{SP}^{2\omega}(\varphi) \propto \chi_{yxy}^2 \cos(\theta)\cos^2(3\varphi)$$

$$I_{SS}^{2\omega}(\varphi) \propto \chi_{yxy}^2 \sin^2(3\varphi)$$

$$I_{\text{Crossed}}^{2\omega}(\varphi) \propto \chi_{yxy}^2 \cos^2(3\phi)$$

$$I_{\text{Parallel}}^{2\omega}(\varphi) \propto \chi_{yxy}^2 \sin^2(3\varphi)$$

The simulated result in the *SP* polarization channel for the *c*-type surface ED contribution is compared with the room temperature data in Fig. S3.

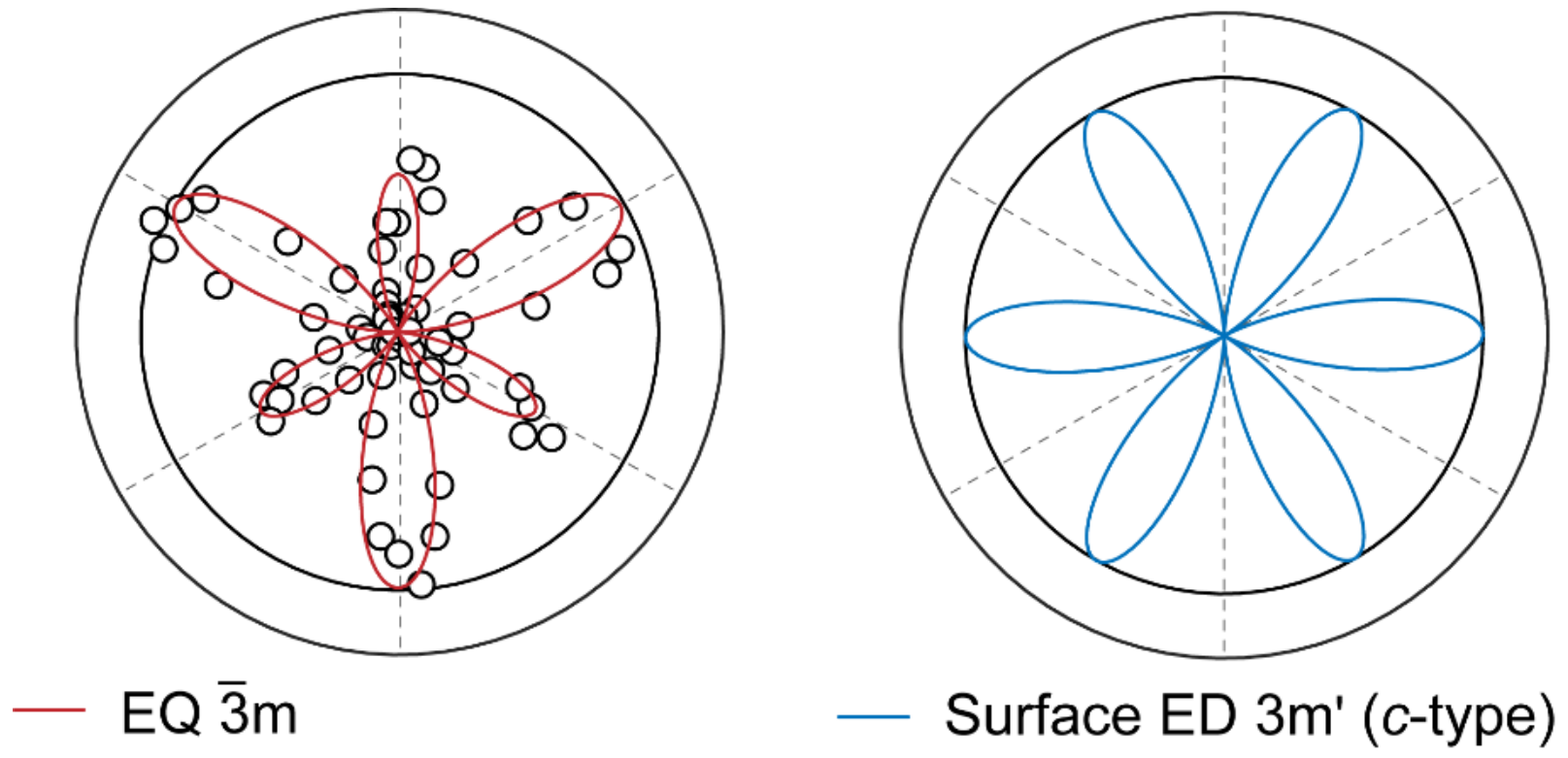

**Figure S3.** SHG RA pattern measured under the *SP* polarization geometry. The data is fitted with the EQ contribution (red) under the $\bar{3}m$ point group. SHG RA pattern simulated for the *c*-type surface ED contribution (blue) under the $3m'$ point group.

**MD SHG of $\bar{3}m$ and $\bar{3}m'$ (*i*-type):**

Since the MD contribution is axial, the SHG response from bulk MD contribution is allowed in centrosymmetric media. The derivation of the MD susceptibility tensor elements is obtained by using[1]

$$\chi_{ijk} = |\sigma|\sigma_{ip}\sigma_{jq}\sigma_{kr}\chi_{pqr}$$

where $\sigma$ is a generating matrix for symmetry operations and $|\sigma|$ is a determinant of $\sigma$ (1 or -1). Non-zero MD tensor elements under $\bar{3}m$ and $\bar{3}m'$ (time-reversal invariant *i*-type) are given as below.

$$\chi_{ijk}^{MD} = \begin{pmatrix} \begin{pmatrix} yxy \\ 0 \\ 0 \end{pmatrix} & \begin{pmatrix} 0 \\ -yxy \\ yxz \end{pmatrix} & \begin{pmatrix} 0 \\ yzx \\ 0 \end{pmatrix} \\ \begin{pmatrix} 0 \\ -yxy \\ -yxz \end{pmatrix} & \begin{pmatrix} -yxy \\ 0 \\ 0 \end{pmatrix} & \begin{pmatrix} -yzx \\ 0 \\ 0 \end{pmatrix} \\ \begin{pmatrix} 0 \\ zyx \\ 0 \end{pmatrix} & \begin{pmatrix} -zyx \\ 0 \\ 0 \end{pmatrix} & \begin{pmatrix} 0 \\ 0 \\ 0 \end{pmatrix} \end{pmatrix}$$

The intensity of SHG RA from the MD contribution can be obtained using an equation[4]

$$I^{2\omega}(\varphi) \propto \left|\hat{e}_{i'}(2\omega)\epsilon_{i'j'k'}\hat{\partial}_{j'}\chi^{MD}_{k'l'm'}(\varphi)\hat{e}_{l'}(\omega)\hat{e}_{m'}(\omega)\right|^2 I(\omega)^2$$

where $\epsilon$ is the Levi-Civita matrix and $\chi^{MD}_{k'l'm'}(\varphi)$ is the transformed matrix into the rotated frame from the tensor elements in the unrotated frame $\chi^{MD}_{ijk}$. The functional forms for all probed polarization geometries are given below.

$$I^{2\omega}_{PP}(\varphi) \propto \cos^2(\theta)\,[\cos^4(\theta) + \sin^4(\theta)][1 + \sin(2\theta)]\left[(\chi_{yzx} + \chi_{zyx})\sin(\theta) + \chi_{yxy}\cos(\theta)\sin(3\varphi)\right]^2$$

$$I^{2\omega}_{PS}(\varphi) \propto \chi^2_{yxy}\cos^6(\theta)\,[1 + \sin(2\theta)]\cos^2(3\varphi)$$

$$I^{2\omega}_{SP}(\varphi) \propto \frac{1}{4}\chi^2_{yxy}[3 + \cos(4\theta)][1 + \sin(2\theta)]\sin^2(3\varphi)$$

$$I^{2\omega}_{SS}(\varphi) \propto \chi^2_{yxy}\cos^2(\theta)\,[1 + \sin(2\theta)]\cos^2(3\varphi)$$

$$I^{2\omega}_{\mathrm{Crossed}}(\phi) \propto \chi^2_{yxy}\sin^2(3\varphi)$$

$$I^{2\omega}_{\mathrm{Parallel}}(\varphi) \propto \chi^2_{yxy}\cos^2(3\varphi)$$

The derived functional form for the *SP* channel predicts the even six lobes of the SHG RA (Fig. S4) as similar in the surface ED contribution under $3m$.

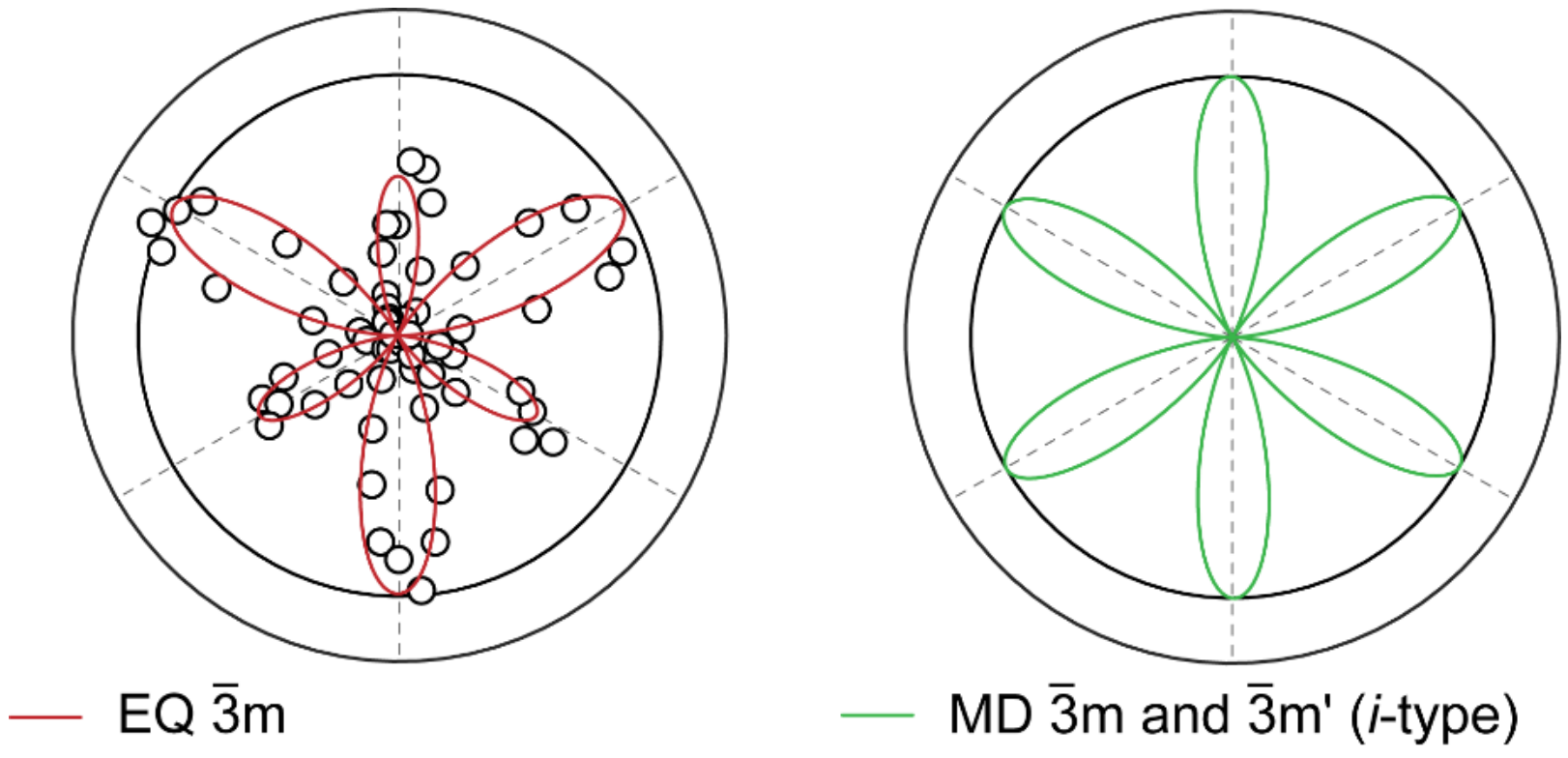

**Figure S4.** SHG RA data for the *SP* channel fitted with the simulated *SP* channel of EQ SHG (red) and SHG RA pattern simulated for MD SHG (green) for $\bar{3}m$ and $\bar{3}m'$ (*i*-type) point groups

**MD SHG of $\bar{3}m'$ (*c*-type):**

For the time-reversal noninvariant (broken) *c*-type MD contribution, the susceptibility tensor elements can be obtained using[1]

$$\chi_{ijk} = (-1)|\sigma|\sigma_{ip}\sigma_{jq}\sigma_{kr}\chi_{pqr}$$

Here, the equation is only applicable for a generating matrix $\sigma$ for operations including the time-reversal operator such as $m'$ in $\bar{3}m'$. Other generating matrixes should be applied following the equation for the *i*-type MD SHG. The obtained susceptibility tensor elements are

$$\chi_{ijk}^{MD} = \begin{pmatrix} \begin{pmatrix} 0 \\ -xyx \\ yyz \end{pmatrix} & \begin{pmatrix} -xyx \\ 0 \\ 0 \end{pmatrix} & \begin{pmatrix} yzy \\ 0 \\ 0 \end{pmatrix} \\ \begin{pmatrix} -xyx \\ 0 \\ 0 \end{pmatrix} & \begin{pmatrix} 0 \\ xyx \\ yyz \end{pmatrix} & \begin{pmatrix} 0 \\ yzy \\ 0 \end{pmatrix} \\ \begin{pmatrix} zyy \\ 0 \\ 0 \end{pmatrix} & \begin{pmatrix} 0 \\ zyy \\ 0 \end{pmatrix} & \begin{pmatrix} 0 \\ 0 \\ zzz \end{pmatrix} \end{pmatrix}$$

The same procedure as the functional forms of the *i*-type MD contribution can be used to obtain the functional forms as given below.

$$I_{PP}^{2\omega}(\varphi) \propto \frac{1}{4}\chi_{xyx}^2 \cos^4(\theta)\,[3+\cos(4\theta)][1+\sin(2\theta)]\cos^2(3\varphi)$$

$$I_{PS}^{2\omega}(\varphi) \propto [1+\sin(2\theta)]\big[\big(\chi_{yyz}+\chi_{yzy}+\chi_{zyy}\big)\cos^2(\theta)\sin(\theta)+\chi_{zzz}\sin^3(\theta) + \chi_{xyx}\cos^3(\theta)\sin(3\varphi)\big]^2$$

$$I_{SP}^{2\omega}(\varphi) \propto \frac{1}{4}\chi_{xyx}^2[3+\cos(4\theta)][1+\sin(2\theta)]\cos^2(3\varphi)$$

$$I_{SS}^{2\omega}(\varphi) \propto [1+\sin(2\theta)]\big[\chi_{zyy}\sin(\theta)-\chi_{xyx}\cos(\theta)\sin(3\varphi)\big]^2$$

$$I_{\mathrm{Crossed}}^{2\omega}(\varphi) \propto \chi_{xyx}^2\cos^2(3\phi)$$

$$I_{\mathrm{Parallel}}^{2\omega}(\varphi) \propto \chi_{xyx}^2\sin^2(3\varphi)$$

The simulated result in the *SP* polarization channel is shown in Fig. S5, excluding the *c*-type MD contribution.

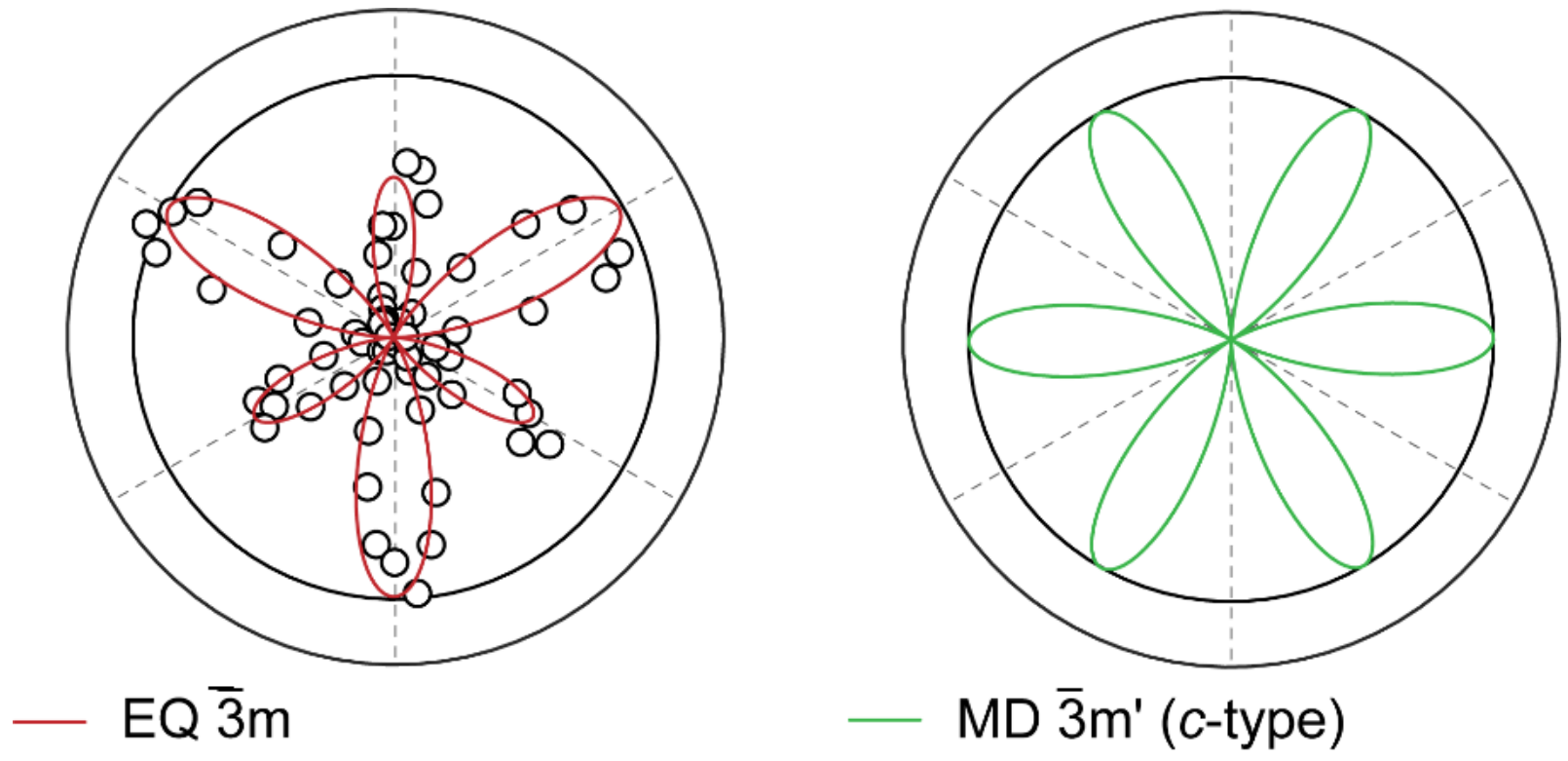

**Figure S5.** SHG RA pattern measured under the *SP* polarization geometry. The data is fitted with the EQ contribution (red) under the $\bar{3}m$ point group. SHG RA pattern simulated for the *c*-type MD contribution (light blue) under the $\bar{3}m'$ point group. The SHG RA pattern from the *c*-type MD contribution shows peak lobe intensities at different $\varphi$ values.

**Electric-field-induced second harmonic generation (EFISH) from the $\bar{3}m$ and $\bar{3}m'$ *i*-type contributions**

The existence of electric field in centrosymmetric media can effectively break the inversion symmetry, resulting in non-zero second harmonic generation. We considered the centrosymmetric media under a built-in electric field along the out-of-plane direction, and thus the nonlinear effective polarization *P* is expressed by

$$P_i(2\omega) \propto \chi_{ijkl}^{EFISH} E_j(\omega)\mathcal{E}_k E_l(\omega)$$

where $\chi^{EFISH}$ is the third order nonlinear optical susceptibility tensor of the electric dipolar contribution under the point group $\bar{3}m$, and $\mathcal{E}$ is the electric field along the out-of-plane direction. $\chi_{ijkl}^{EFISH}$ for $\bar{3}m$ and $\bar{3}m'$ *i*-type contributions shares the same tensor elements as bulk $\chi_{ijkl}^{EQ}$ for $\bar{3}m$ and $\bar{3}m'$ *i*-type contributions with 11 nonzero tensor elements as given below.

$$\chi_{ijkl}^{EFISH} = \begin{pmatrix} \begin{pmatrix} xxxx & 0 & 0 \\ 0 & xxyy & -yyyz \\ 0 & -yyzy & xxzz \end{pmatrix} & \begin{pmatrix} 0 & xyxy & -yyyz \\ xxyy & 0 & 0 \\ -yyzy & 0 & 0 \end{pmatrix} & \begin{pmatrix} 0 & -yyyz & xzxz \\ -yyyz & 0 & 0 \\ xxzz & 0 & 0 \end{pmatrix} \\ \begin{pmatrix} 0 & xxyy & -yyyz \\ xyxy & 0 & 0 \\ -yyzy & 0 & 0 \end{pmatrix} & \begin{pmatrix} xxyy & 0 & 0 \\ 0 & xxxx & yyyz \\ 0 & yyzy & xxzz \end{pmatrix} & \begin{pmatrix} -yyyz & 0 & 0 \\ 0 & yyyz & xzxz \\ 0 & xxzz & 0 \end{pmatrix} \\ \begin{pmatrix} 0 & -zyyy & zzxx \\ -zzyy & 0 & 0 \\ zxzx & 0 & 0 \end{pmatrix} & \begin{pmatrix} -zyyy & 0 & 0 \\ 0 & zyyy & zzxx \\ 0 & zxzx & 0 \end{pmatrix} & \begin{pmatrix} zzxx & 0 & 0 \\ 0 & zzxx & 0 \\ 0 & 0 & zzzz \end{pmatrix} \end{pmatrix}$$

The susceptibility matrix was transformed into the rotated frame, and then the EFISH intensity is given from

$$I^{2\omega}(\varphi) \propto \left| \hat{e}_{i'}(2\omega)\chi_{i'j'k'l'}^{EFISH}(\varphi)\hat{e}_{j'}(\omega)\mathcal{E}_{k'}\hat{e}_{l'}(\omega) \right|^2 I(\omega)^2$$

The polarization of the incident and reflected light was chosen to be consistent with our experimental polarization geometry, and the functional forms under selected polarization geometries are given below.

$$I_{PP}^{2\omega}(\varphi) \propto [\chi_{zxzx}\cos(\theta)^2\sin(\theta) + \chi_{zzzz}\sin^3(\theta)]^2 + \cos^4(\theta)\left[2\chi_{xxzz}\sin(\theta) + \chi_{yyzy}\cos(\theta)\sin^2(3\varphi)\right]^2$$

$$I_{PS}^{2\omega}(\varphi) \propto \chi_{yyzy}^2\cos^4(\theta)\cos^2(3\varphi)$$

$$I_{SP}^{2\omega}(\varphi) \propto \chi_{zxzx}^2\sin^2(\theta) + \chi_{yyzy}^2\cos^2(\theta)\sin^2(3\varphi)$$

$$I_{SS}^{2\omega}(\varphi) \propto \chi_{yyzy}^2\cos^2(3\varphi)$$

$$I_{\mathrm{Crossed}}^{2\omega}(\varphi) \propto \chi_{yyzy}^2\sin^2(3\varphi)$$

$$I_{\mathrm{Parallel}}^{2\omega}(\varphi) \propto \chi_{yyzy}^2\cos^2(3\varphi)$$

In the *SP* polarization channel, the simulated SHG RA pattern preserves six-fold rotational symmetry with six even lobes, which is inconsistent with three-fold rotation symmetry of EQ $\bar{3}m$ in the *SP* channel given in Fig. S6.

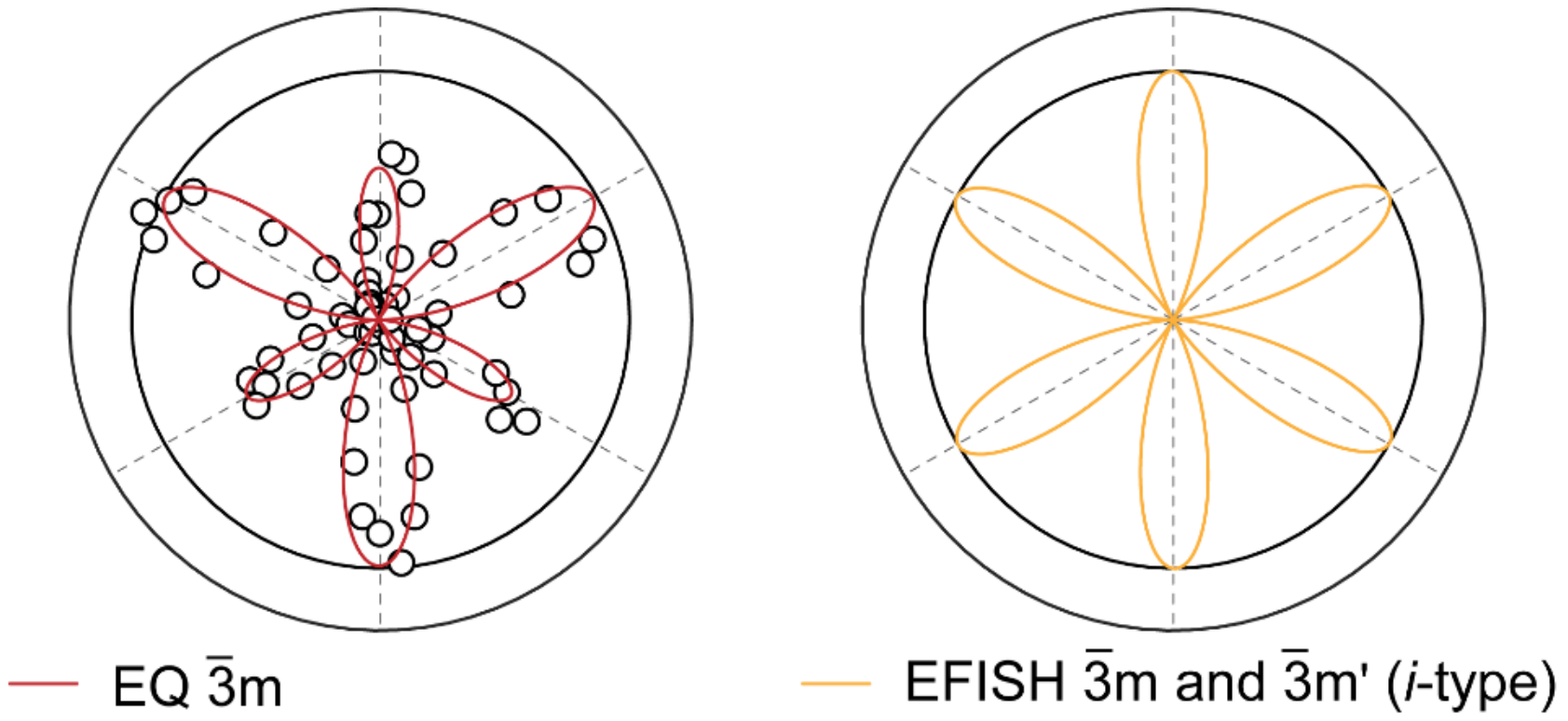

**Figure S6.** SHG RA data for the *SP* channel fitted with the simulated *SP* channel of EQ SHG (red) and SHG RA pattern simulated for EFISH (yellow) from the *i*-type $\bar{3}m$ and $\bar{3}m'$ contributions

**EFISH from the *c*-type contribution under the $\bar{3}m'$ magnetic point group**

The time-reversal broken *c*-type tensor elements for the EFISH can be obtained using a following equation[1]

$$\chi_{ijkl} = (-1)\sigma_{ip}\sigma_{jq}\sigma_{kr}\sigma_{ls}\chi_{pqrs}$$

Here, the multiplication of -1 is only applied for the generating matrix for operations including the time-reversal operator. The obtained susceptibility tensor elements for the *c*-type EFISH for $\bar{3}m'$ are given below

$$\chi_{ijkl}^{\mathrm{EFISH}} = \begin{pmatrix} \begin{pmatrix} 0 & xxxy & xxxz \\ -yyxy & 0 & 0 \\ xxzx & 0 & 0 \end{pmatrix} & \begin{pmatrix} -yxyy & 0 & 0 \\ 0 & xyyy & -xxxz \\ 0 & -xxzx & xyzz \end{pmatrix} & \begin{pmatrix} xxxz & 0 & 0 \\ 0 & -xxxz & xzyz \\ 0 & xyzz & 0 \end{pmatrix} \\ \begin{pmatrix} -xyyy & 0 & 0 \\ 0 & yxyy & -xxxz \\ 0 & -xxzx & -xyzz \end{pmatrix} & \begin{pmatrix} 0 & yyxy & -xxxz \\ -xxxy & 0 & 0 \\ -xxzx & 0 & 0 \end{pmatrix} & \begin{pmatrix} 0 & -xxxz & -xzyz \\ -xxxz & 0 & 0 \\ -xyzz & 0 & 0 \end{pmatrix} \\ \begin{pmatrix} zxxx & 0 & 0 \\ 0 & -zxxx & zxyz \\ 0 & zxzy & 0 \end{pmatrix} & \begin{pmatrix} 0 & -zxxx & -zxyz \\ -zxxx & 0 & 0 \\ -zxzy & 0 & 0 \end{pmatrix} & \begin{pmatrix} 0 & zzxy & 0 \\ -zzxy & 0 & 0 \\ 0 & 0 & 0 \end{pmatrix} \end{pmatrix}$$

The susceptibility matrix was transformed into the rotated frame, and then the functional forms of EFISH intensity under employed polarization geometries are derived as below.

$$I_{PP}^{2\omega}(\varphi) \propto \chi_{xxzx}^{2} \cos^{6}(\theta) \cos^{2}(3\varphi)$$

$$I_{PS}^{2\omega}(\varphi) \propto \left[-2\chi_{xyzz} \sin(\theta) \cos(\theta) + \chi_{xxzx} \cos^{2}(\theta) \sin(3\varphi)\right]^{2}$$

$$I_{SP}^{2\omega}(\varphi) \propto \chi_{xxzx}^{2} \cos^{2}(\theta) \cos^{2}(3\varphi)$$

$$I_{SS}^{2\omega}(\varphi) \propto \chi_{xxzx}^{2}\, sin^{2}(3\varphi)$$

$$I_{\mathrm{Crossed}}^{2\omega}(\varphi) \propto \chi_{xxzx}^{2} \cos^{2}(3\varphi)$$

$$I_{\mathrm{Parallel}}^{2\omega}(\varphi) \propto \chi_{xxzx}^{2} \sin^{2}(3\varphi)$$

The simulated SHG RA pattern in the *SP* polarization channel exhibits a distinct angular dependence with the preserved six-fold rotational symmetry from the data obtained in the *SP* channel as shown in Fig. S7.

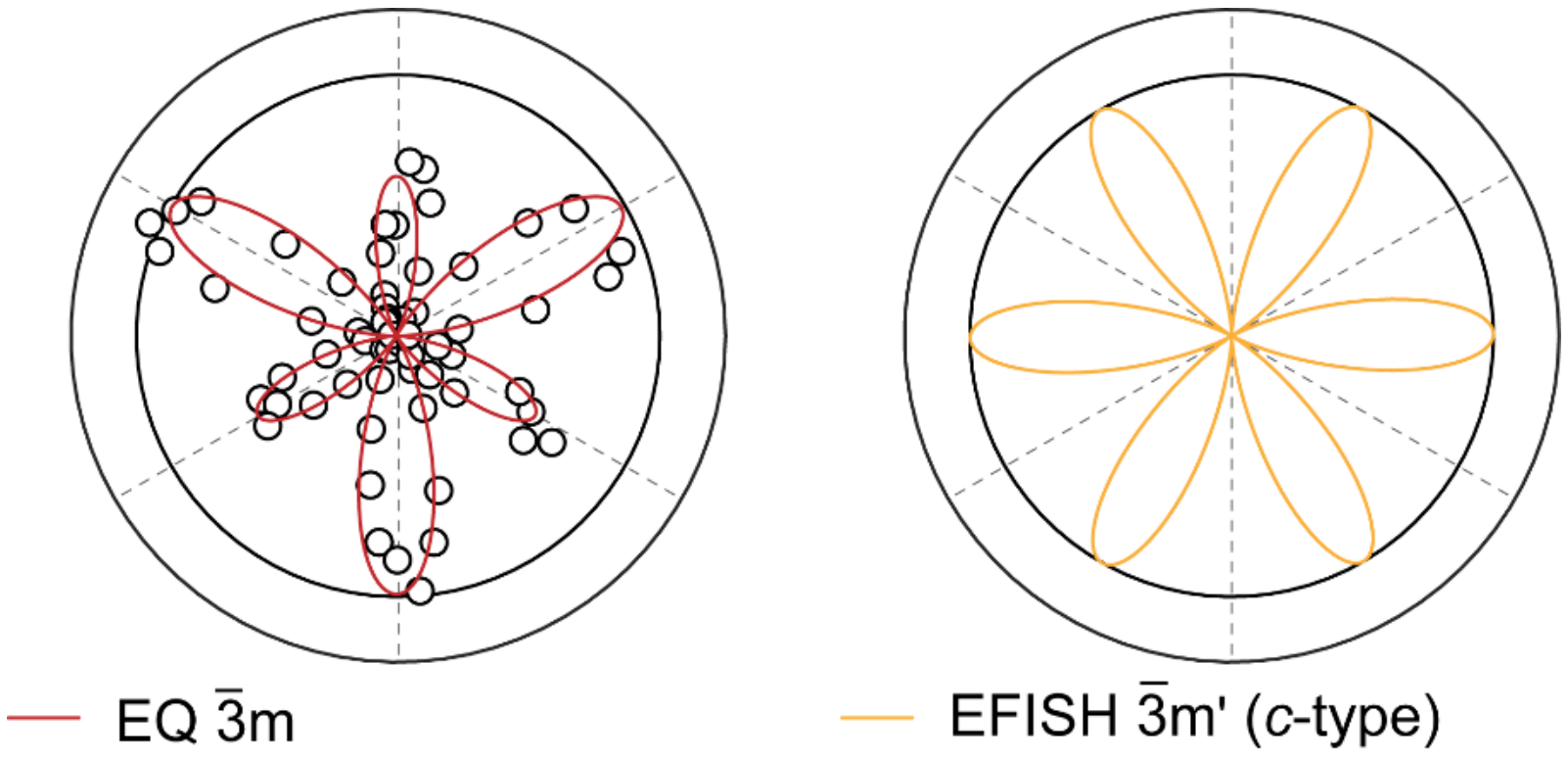

**Figure S7.** SHG RA data for the *SP* channel fitted with the simulated *SP* channel of EQ SHG (red) and SHG RA pattern simulated for EFISH (yellow) from the *c*-type $\bar{3}m'$ contribution

## Section 3. SHG RA data in the Cartesian coordinate for the RA patterns

In Figure 3 in the main text, the Cartesian plots of fitted SHG RA profiles are provided to visualize the rotational phase shift $\Delta\varphi$ at low temperatures. In this section, we provide zoom-in Cartesian plots with both raw data and the fitted results to better illustrate the mirror symmetry breaking manifested by the rotational phase shift. Figure S8 shows the normalized SHG RA data (squares and dots) and their fits (solid curves) in the crossed (Fig. S8a) and the parallel (Fig. S8c) channels both at 200 K (red) and 90 K (blue) that are above and below the magnetic phase transitions, respectively. The zoomed-in Cartesian plots at the two temperatures (Figs. S8b and S8d) clearly show that the 90 K data points are consistently shifted towards the higher angles from the 200 K ones, and that their fits also exhibit a rotational phase shift ($\Delta\varphi$).

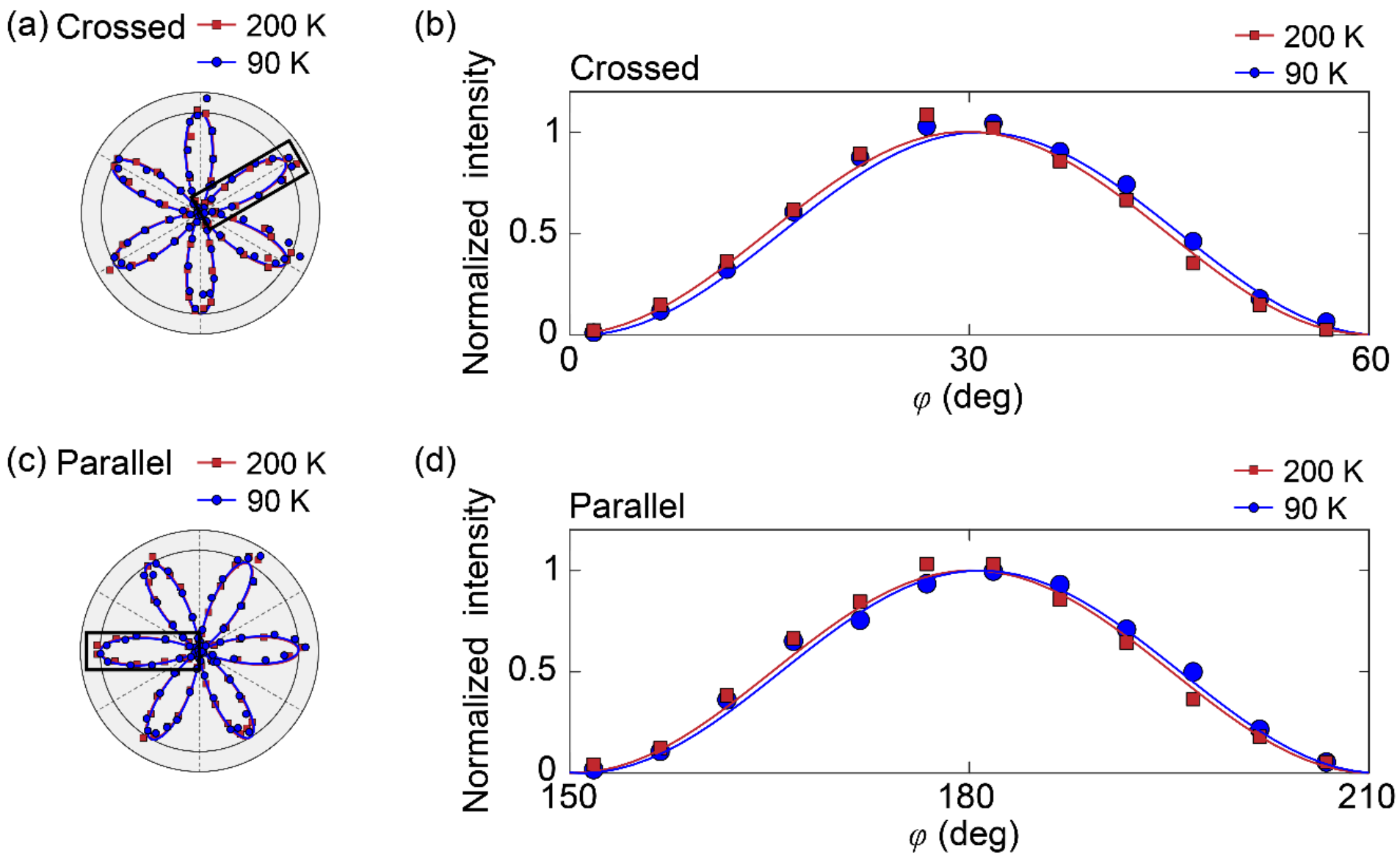

**Fig. S8** SHG RA patterns for (a) crossed and (c) parallel channels measured at 200 K (red) and 90 K (blue) in the polar coordinate. The Cartesian plots in (b) the crossed and (d) the parallel channels include the data enclosed by the rectangles in (a) and (c).

We also display the Cartesian plots of raw data and fitted curves measured at 200 K and 90 K in *SS*/*SP*/*PS*/*PP* polarization channels under the oblique incident geometry in Fig. S9. The results of all polarization channels consistently show that the RA patterns shift towards higher angles at 90 K.

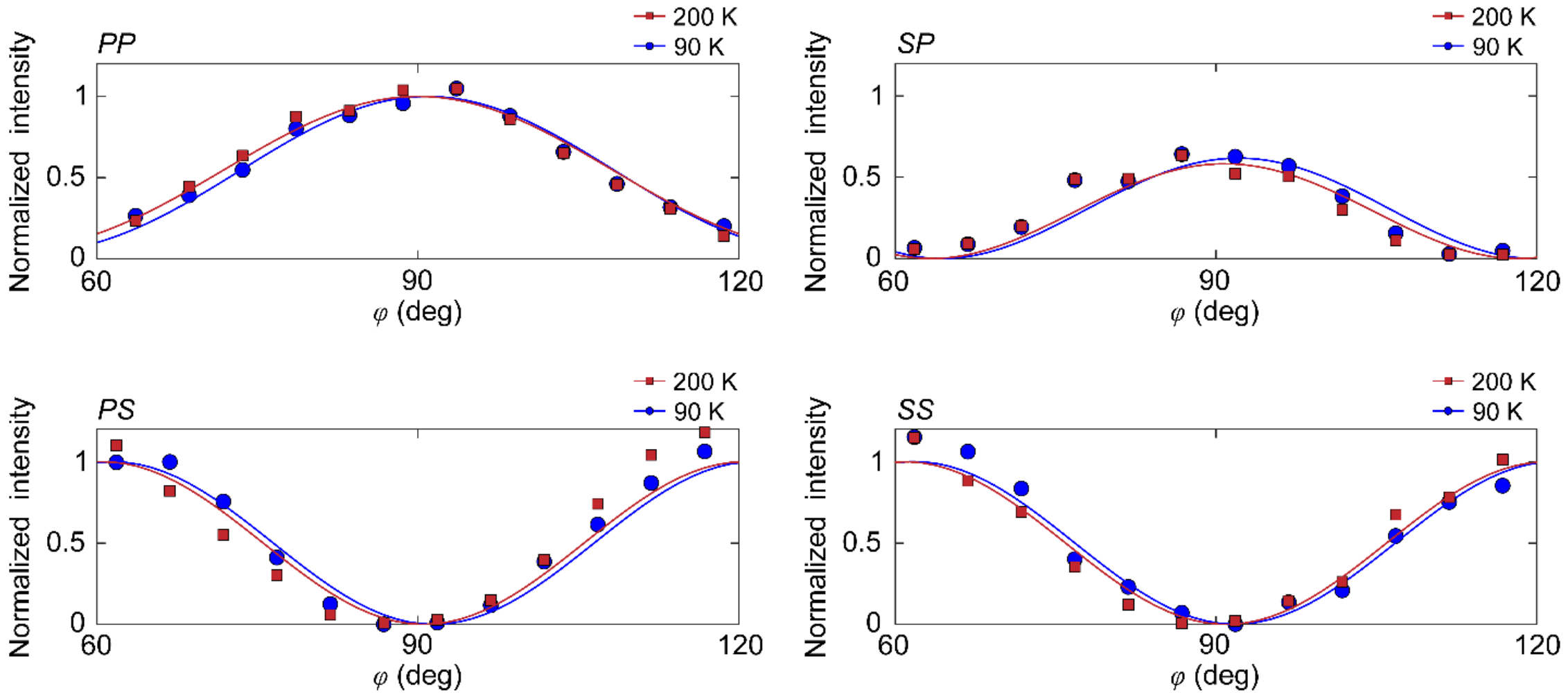

**Fig. S9** Cartesian plots of SHG RA patterns for *PP*/*PS*/*SP*/*SS* polarization channels measured at 200 K (red) and 90 K (blue)

**Section 4. Involved SHG processes, their coherent superpositions to fit the data, and their relationship to the magnetic order parameters**

From temperature-dependent SHG data, we observed two SHG anomalies (intensity enhancement and rotation of SHG RA patterns) at two different temperatures ($T_{C,1}$ and $T_{C,2}$). In order to understand these anomalies, emerging SHG radiation sources upon the phase transitions have to be identified. In this section, first we introduce the derivation of functional forms of EQ SHG intensity from $c$-type contribution for $\bar{3}m'$ magnetic point group, which leads to the rotation of SHG RA patterns. Second, with the derived functional forms, we show the coherent interference between crystallographic, $i$-type, and $c$-type contributions for three temperature ranges, which explains the intensity enhancement and the rotation of SHG RA patterns as well as their temperature dependence in a consistent manner. Finally, we interpret the temperature-dependent SHG data with respect to magnetic order parameters so that the characteristics of the phase transitions and ordered states can be deduced.

**Functional forms of EQ SHG of *c*-type $\bar{3}m'$**

The time-reversal broken $c$-type tensor elements for the polar EQ SHG can be obtained using a following equation[1]

$$\chi_{ijkl} = (-1)\sigma_{ip}\sigma_{jq}\sigma_{kr}\sigma_{ls}\chi_{pqrs}$$

Here, the multiplication of -1 is only applied for the generating matrix for operations including the time-reversal operator. The obtained susceptibility tensor elements for the $c$-type EQ SHG for $\bar{3}m'$ are given below

$$\chi_{ijkl}^{EQ\,(c)} = \begin{pmatrix} \begin{pmatrix} 0 & xxxy & xxxz \\ -yyxy & 0 & 0 \\ xxzx & 0 & 0 \end{pmatrix} & \begin{pmatrix} -yxyy & 0 & 0 \\ 0 & xyyy & -xxxz \\ 0 & -xxzx & xyzz \end{pmatrix} & \begin{pmatrix} xxxz & 0 & 0 \\ 0 & -xxxz & xzyz \\ 0 & xyzz & 0 \end{pmatrix} \\ \begin{pmatrix} -xyyy & 0 & 0 \\ 0 & yxyy & -xxxz \\ 0 & -xxzx & -xyzz \end{pmatrix} & \begin{pmatrix} 0 & yyxy & -xxxz \\ -xxxy & 0 & 0 \\ -xxzx & 0 & 0 \end{pmatrix} & \begin{pmatrix} 0 & -xxxz & -xzyz \\ -xxxz & 0 & 0 \\ -xyzz & 0 & 0 \end{pmatrix} \\ \begin{pmatrix} zxxx & 0 & 0 \\ 0 & -zxxx & zxyz \\ 0 & zxzy & 0 \end{pmatrix} & \begin{pmatrix} 0 & -zxxx & -zxyz \\ -zxxx & 0 & 0 \\ -zxzy & 0 & 0 \end{pmatrix} & \begin{pmatrix} 0 & zzxy & 0 \\ -zzxy & 0 & 0 \\ 0 & 0 & 0 \end{pmatrix} \end{pmatrix}$$

With the obtained tensor elements, the functional forms under parallel and crossed polarization geometries at the normal incidence are

$$I_{\text{Crossed}}^{2\omega}(\varphi) \propto \chi_{xxzx}^{2}\cos^{2}(3\varphi)$$

$$I_{\text{Parallel}}^{2\omega}(\varphi) \propto \chi_{xxzx}^{2}\sin^{2}(3\varphi)$$

which exhibits distinct $\varphi$-dependence from those for crystallographic and $i$-type contributions under $\bar{3}m$ and $\bar{3}m'$ as given in section 1. Due to this angular difference, the emerging $c$-type contribution across the phase transitions can lead to the rotation of the SHG RA patterns, which is described as follows.

**Temperature dependence of SHG radiation sources and their coherent superposition**

From the experiment, we observed two phase transitions manifested by the anomalies of the SHG RA patterns at $T_{C,1}$ = 175 K and $T_{C,2}$ = 120 K. In order to understand characteristics of these two phase transitions, we identify and analyze the emergences and anomalies of the temperature dependent SHG RA patterns for three different temperature ranges, attributed to emerging nonlinear susceptibilities and relevant magnetic order parameters, which are summarized in Table 1 given below.

*Table 1. Involved nonlinear susceptibility and order parameters at different temperatures.*

| **Temperature** | $T$ > 175 K | 175 K > $T$ > 120 K | 120 K > $T$ |
|---|---|---|---|
| **Contributing susceptibilities** | $\chi_{\mathrm{S}}^{\mathrm{EQ}}$ | $\chi_{\mathrm{S}}^{\mathrm{EQ}}, \chi_{\mathrm{M_I}}^{\mathrm{EQ}\,(i)}, \chi_{\mathrm{M_I}}^{\mathrm{EQ}\,(c)}$ | $\chi_{\mathrm{S}}^{\mathrm{EQ}}, \chi_{\mathrm{M_I}}^{\mathrm{EQ}\,(i)}, \chi_{\mathrm{M_I}}^{\mathrm{EQ}\,(c)}, \chi_{\mathrm{M_{II}}}^{\mathrm{EQ}\,(i)}, \chi_{\mathrm{M_{II}}}^{\mathrm{EQ}\,(c)}$ |
| **Order parameters** | NA | $\mathrm{M_I} \neq 0, \mathrm{M_{II}} = 0$ | $\mathrm{M_I} \neq 0, \mathrm{M_{II}} \neq 0$ |

Here, $\chi_{\mathrm{S}}^{\mathrm{EQ}}$ is responsible for the crystallographic contribution above 175 K. $\chi_{\mathrm{M_I}}^{\mathrm{EQ}\,(i)}$ and $\chi_{\mathrm{M_{II}}}^{\mathrm{EQ}\,(i)}$ are the time-invariant $i$-type tensors contributing from the magnetic states with order parameters $\mathrm{M_I}$ and $\mathrm{M_{II}}$ below 175 K and 120 K, respectively. Due to their time-invariant nature, $\chi_{\mathrm{M_I}}^{\mathrm{EQ}\,(i)}$ and $\chi_{\mathrm{M_{II}}}^{\mathrm{EQ}\,(i)}$ share the tensor form identical to crystallographic $\chi_{\mathrm{S}}^{\mathrm{EQ}}$, and are proportional to the even powers of the order parameters. In contrast, $c$-type $\chi_{\mathrm{M_I}}^{\mathrm{EQ}\,(c)}$ and $\chi_{\mathrm{M_{II}}}^{\mathrm{EQ}\,(c)}$ are time-noninvariant, and thus have different tensor components from $i$-type counterparts, so that the emergence of $c$-type tensors is responsible for the rotation of SHG RA patterns as described in section 3. These $c$-type tensors scale with the odd powers of the order parameters. Below we provide how the functional forms of intensity and rotation of the SHG RA patterns are derived with respect to crystallographic ($\chi_{\mathrm{S}}^{\mathrm{EQ}}$), magnetic $i$-type ($\chi_{\mathrm{M_I}}^{\mathrm{EQ}\,(i)}$, $\chi_{\mathrm{M_{II}}}^{\mathrm{EQ}\,(i)}$) and $c$-type ($\chi_{\mathrm{M_I}}^{\mathrm{EQ}\,(c)}$, $\chi_{\mathrm{M_{II}}}^{\mathrm{EQ}\,(c)}$) nonlinear susceptibilities as well as magnetic order parameters ($\mathrm{M_I}$, $\mathrm{M_{II}}$) for three temperature ranges with a focus of crossed polarization channel.

1) Above $T_{C,1}$, the SHG intensity in the crossed polarization channel as derived in section 1 is given by

$$I_{\mathrm{Crossed}}^{2\omega}(\varphi) \propto \left|\chi_{yyzy}^{\mathrm{EQ}\,(s)} \sin(3\varphi)\right|^2$$

2) Between $T_{C,2} < T < T_{C,1}$ where $\mathrm{M_I} \neq 0$ and $\mathrm{M_{II}} = 0$, EQ SHG contributions from $\mathrm{M_I}$ coherently interfere with the crystallographic contribution, and the functional form in the crossed polarization channel becomes

$$I_{\mathrm{Crossed}}^{2\omega}(\varphi) \propto \left|\chi_{yyzy}^{\mathrm{EQ}\,(s)} \sin(3\varphi) + \chi_{yyzy}^{\mathrm{EQ}\,(i,\mathrm{M_I})} \sin(3\varphi) - \chi_{xxzx}^{\mathrm{EQ}\,(c,\mathrm{M_I})} \cos(3\varphi)\right|^2$$

This form can be further simplified using the trigonometric function properties so that

$$I_{\text{Crossed}}^{2\omega}(\varphi) \propto I_0 \sin^2[3(\phi - \Delta\phi)],$$

where $I_0 = \left(\chi_{yyzy}^{\text{EQ}\,(s)} + \chi_{yyzy}^{\text{EQ}\,(i,\text{M}_\text{I})}\right)^2 + \left(\chi_{xxzx}^{\text{EQ}\,(c,\text{M}_\text{I})}\right)^2$ and $\Delta\varphi = \frac{1}{3}\tan^{-1}\left(\frac{\chi_{xxzx}^{\text{EQ}\,(c,\text{M}_\text{I})}}{\chi_{yyzy}^{\text{EQ}\,(s)} + \chi_{yyzy}^{\text{EQ}\,(i,\text{M}_\text{I})}}\right)$. We note that in our experimental data, the $\Delta\varphi$ changes by ~ 1° whereas $I_0$ increases by a factor of ~1.5. This observation assures that $\chi_{xxzx}^{\text{EQ}\,(c,\text{M}_\text{I})}$ is nearly 20 times smaller than $\chi_{yyzy}^{\text{EQ}\,(s)} + \chi_{yyzy}^{\text{EQ}\,(i,\text{M}_\text{I})}$, and thus that $\left(\chi_{xxzx}^{\text{EQ}\,(c,\text{M}_\text{I})}\right)^2$ is more than two orders of magnitude smaller than $\left(\chi_{yyzy}^{\text{EQ}\,(s)} + \chi_{yyzy}^{\text{EQ}\,(i,\text{M}_\text{I})}\right)^2$ .Therefore, $I_0 \approx \left(\chi_{yyzy}^{\text{EQ}\,(s)} + \chi_{yyzy}^{\text{EQ}\,(i,\text{M}_\text{I})}\right)^2$ and $\Delta\varphi \propto \chi_{xxzx}^{\text{EQ}\,(c,\text{M}_\text{I})}$.

3) Below $T_{\text{C},2}$ where $\text{M}_\text{I} \neq 0$ and $\text{M}_\text{II} \neq 0$, the coherent interference of EQ SHG radiation sources from crystallographic, $\text{M}_\text{I}$ and $\text{M}_\text{II}$ contributions result in the functional form in the crossed polarization channel to be

$$I_{\text{Crossed}}^{2\omega}(\varphi) \propto \left|\left(\chi_{yyzy}^{\text{EQ}\,(s)} + \chi_{yyzy}^{\text{EQ}\,(i,\text{M}_\text{I})} + \chi_{yyzy}^{\text{EQ}\,(i,\text{M}_\text{II})}\right)\sin(3\varphi) - \left(\chi_{xxzx}^{\text{EQ}\,(c,\text{M}_\text{I})} + \chi_{xxzx}^{\text{EQ}\,(c,\text{M}_\text{II})}\right)\cos(3\varphi)\right|^2$$

which can be simplified to be

$$I_{\text{Crossed}}^{2\omega}(\varphi) \propto I_0 \sin^2[3(\phi - \Delta\phi)],$$

where $I_0 \approx \left(\chi_{yyzy}^{\text{EQ}\,(s)} + \chi_{yyzy}^{\text{EQ}\,(i,\text{M}_\text{I})} + \chi_{yyzy}^{\text{EQ}\,(i,\text{M}_\text{II})}\right)^2$ and $\Delta\varphi \propto \chi_{xxzx}^{\text{EQ}\,(c,\text{M}_\text{I})} + \chi_{xxzx}^{\text{EQ}\,(c,\text{M}_\text{II})}$.

Then, we analyze temperature dependence of SHG intensity $I_0$ and RA rotation $\Delta\varphi$ with these derived function forms by accounting the magnetic order parameters $\text{M}_\text{I}$ and $\text{M}_\text{II}$. As described above, *i*-type and *c*-type susceptibility tensors are proportional to even and odd powers of the order parameters, respectively. In the consideration of the leading order approximation, we have the following expressions: $\chi_{yyzy}^{\text{EQ}\,(s)} = c$, $\chi_{yyzy}^{\text{EQ}\,(i,\text{M}_\text{I})} = a_\text{I}\text{M}_\text{I}^2$, $\chi_{yyzy}^{\text{EQ}\,(i,\text{M}_\text{II})} = a_\text{II}\text{M}_\text{II}^2$, $\chi_{xxzx}^{\text{EQ}\,(c,\text{M}_\text{I})} = b_\text{I}\text{M}_\text{I}$, and $\chi_{xxzx}^{\text{EQ}\,(c,\text{M}_\text{II})} = b_\text{II}\text{M}_\text{II}$, where $c$, $a_\text{I}$, $a_\text{II}$, $b_\text{I}$, and $b_\text{II}$ are constant. Since the observed magnetic phase transition are the second order, the magnetic order parameters can be expressed by $\text{M}_\text{I} = \alpha_\text{I}|T\text{-}T_{\text{C},1}|^{\beta_{\text{C},1}}$ and $\text{M}_\text{II} = \alpha_\text{II}|T\text{-}T_{\text{C},2}|^{\beta_{\text{C},2}}$ where $\alpha$ is constant and $\beta$ is critical exponent of the phase transitions. By taking all derived functional forms together, $I_0(T)$ and $\Delta\varphi$ have the following relations with the leading order approximation:

*Above $T_{C,1}$,*

$$I_0(T) = c^2 = C,$$

$$\Delta\varphi(T) = 0$$

*Between $T_{C,1}$ and $T_{C,2}$,*

$$I_0(T) \approx \left(c + a_I\text{M}_\text{I}^2\right)^2 \approx c^2 + 2ca_\text{I}\text{M}_\text{I}^2 = C + A_\text{I}|T\text{-}T_{\text{C},1}|^{2\beta_{\text{C},1}},$$

$$\Delta\varphi(T) \propto b_{\mathrm{I}}\mathrm{M}_{\mathrm{I}} = B_{\mathrm{I}}\left|T\text{-}T_{\mathrm{C},1}\right|^{\beta_{\mathrm{C},1}}$$

*Below $T_{C,2}$,*

$$I_0(T) \approx \left(c + a_I\mathrm{M}_{\mathrm{I}}^2 + a_{II}\mathrm{M}_{\mathrm{II}}^2\right)^2 \approx c^2 + 2ca_{\mathrm{I}}\mathrm{M}_{\mathrm{I}}^2 + 2ca_{\mathrm{II}}\mathrm{M}_{\mathrm{II}}^2 = C + A_{\mathrm{I}}\left|T\text{-}T_{\mathrm{C},1}\right|^{2\beta_{\mathrm{C},1}} + A_{\mathrm{II}}\left|T\text{-}T_{\mathrm{C},2}\right|^{2\beta_{\mathrm{C},2}}$$

$$\Delta\varphi(T) \propto b_{\mathrm{I}}\mathrm{M}_{\mathrm{I}} + b_{\mathrm{II}}\mathrm{M}_{\mathrm{II}} = B_{\mathrm{I}}\left|T\text{-}T_{\mathrm{C},1}\right|^{\beta_{\mathrm{C},1}} + B_{\mathrm{II}}\left|T\text{-}T_{\mathrm{C},2}\right|^{\beta_{\mathrm{C},2}}$$

These are the temperature dependent functional forms that we used to fit the data in Figure 4 of the main text. From the fitting results of critical exponents $\beta_{\mathrm{C},1} \approx 0.31$ and $\beta_{\mathrm{C},2} \approx 0.35$ obtained from both intensity and rotation data, the first and second phase transitions at $T_{\mathrm{C},1}$ and $T_{\mathrm{C},2}$ can be characterized by 3-dimensional Ising-type and XY-type, respectively.

## Section 5. Minor contributions from SHG radiation sources other than EQ SHG

In this section, we consider potential magnetism-induced SHG radiation sources such as surface ED, bulk MD, and EFISH contributions emerging upon the magnetic phase transitions. From the experimental data in the *SP* polarization channel, we explicitly observe the three-fold rotation symmetry manifested by alternating lobes as shown in Fig. S10 (a). This feature of the alternating lobes is preserved across the phase transitions down to $T = 90$ K. Conversely, the simulated potential *i*-type and *c*-type SHG radiation sources in Figs. S10 (b) and (c) show six even lobes in the *SP* polarization channel. This result indicates that the alternating lobes have to be shaped to the even six lobes across the phase transitions if the considered potential sources significantly contribute to the SHG signal through the interference with the crystallographic EQ SHG. This can be confirmed by measuring the intensity ratio between large and small lobes as a function of temperature. Figure S10 (d) shows that the intensity ratio barely changes down to 90 K while the intensity increases by nearly a factor of 1.5. This contrast strongly suggests that the magnetism-induced ED, MD, and EFISH contributions have to be very minor even if present.

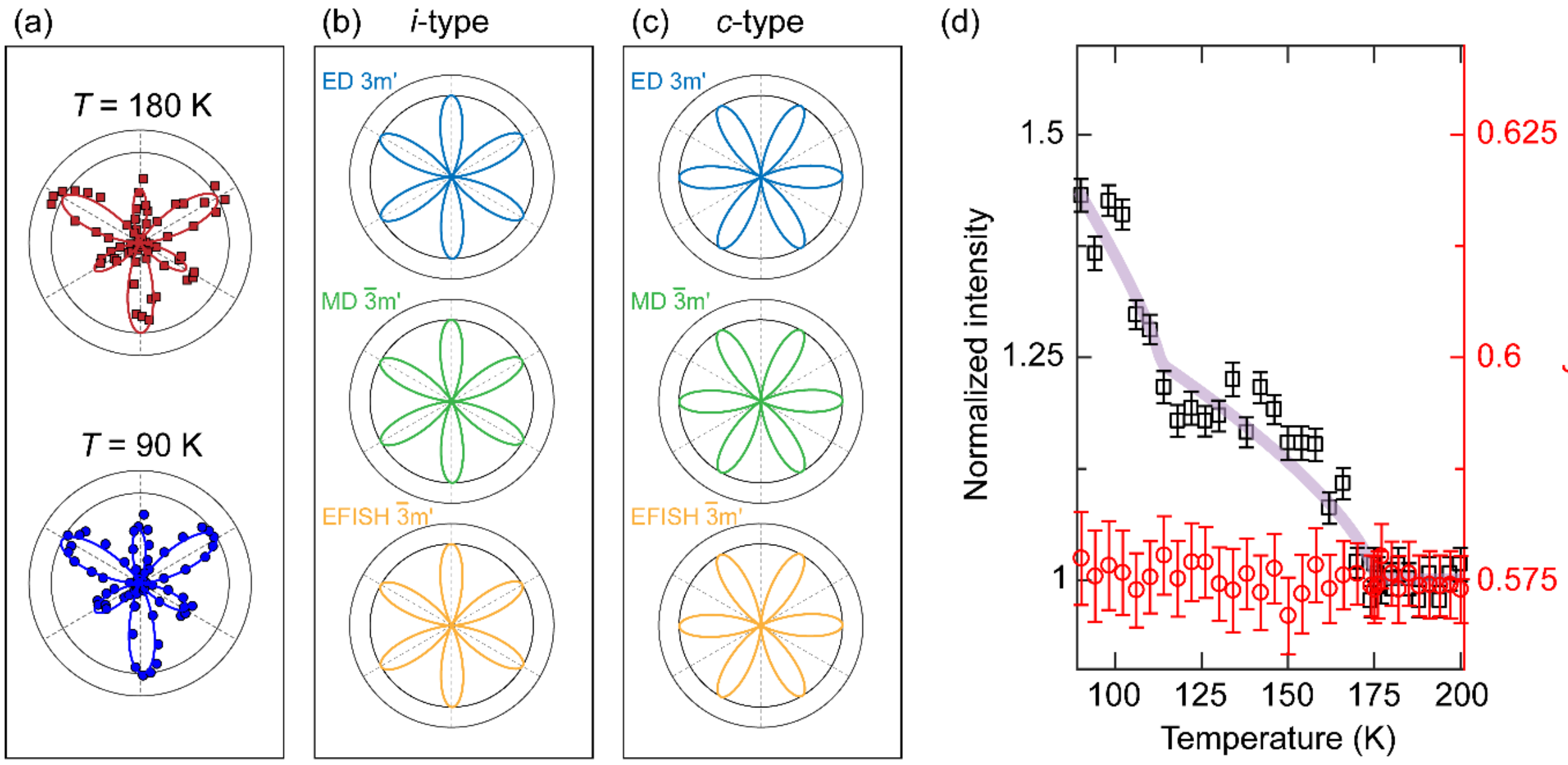

**Figure S10.** (a) SHG RA data for the *SP* channel measured above $T_{C,1}$ ($T = 180$ K) and below $T_{C,2}$ ($T = 90$ K). (b) and (c) The simulated time-invariant *i*-type and time-noninvariant *c*-type radiation sources from magnetism-induced ED ($3m'$), MD ($\bar{3}m'$), and EFISH ($\bar{3}m'$) contributions under the *SP* polarization geometry. (d) Temperature-dependence of the SHG intensity and the intensity ratio between large and small lobes for the *SP* polarization channel.

## Section 6. Comment on symmetry evolution across the phase transitions

In the analysis of temperature-dependent SHG data, we consider that the magnetic point groups for $M_{I}$ and $M_{II}$ are both $\bar{3}m'$, which seemingly suggests no additional symmetry breaking across $T_{C,2}$. If this were to be the case, someone would reasonably raise a question of whether $M_{II}$ is a phase transition or a cross-over phenomenon (see Fig. S11 for the distinction between a phase transition and a cross-over: the cross-over departs from the order-parameter-like onset). We approach this question with two steps below.

a. The crystal structure of $Co_3Sn_2S_2$ obeys the point group $\bar{3}m$, but the kagome magnetic lattice formed by Co ions has the point group $6/mmm$ that is a parent point group of $\bar{3}m$ and has higher symmetries. If we use the Co kagome magnetic lattice to consider the emergence of the two magnetic states, $M_{I}$ and $M_{II}$, we find that (i) across $T_{C,1}$, the point group evolves from $6/mmm$ of the structure to $6/mm'm'$ of the $M_{I}$ phase, breaking the time-reversal, and three vertical and three diagonal mirror symmetries; (ii) across $T_{C,2}$, the point group changes from $6/mm'm'$ of $M_{I}$ to $\bar{3}m'$, breaking the 6-fold rotational symmetry into 3-fold and also losing one $m$ symmetry normal to the 3-fold rotational axis. In this way, there are subsequent broken symmetries at both $T_{C,1}$ and $T_{C,2}$.

b. In this manuscript, we used the crystalline point group of $Co_3Sn_2S_2$, $\bar{3}m$, to perform the symmetry analysis, because our optical SHG RA measurement probes multiple electronic bands, more than the Co kagome lattice. In this way of analysis, both $M_{I}$ and $M_{II}$ take the same magnetic point group, $\bar{3}m'$. Because the symmetries further broken across $T_{C,2}$ from the Co kagome lattice analysis above, e.g., the 6-fold rotational and the one mirror normal to the out-of-plane rotational axis, are already broken by the crystal lattice through the presence of Sn and S ions, one may consider the lattice as an "effective" symmetry-breaking field applied across $T_{C,2}$. Strictly speaking, this presence of the symmetry breaking field for $M_{II}$ makes it a cross-over rather than a phase transition. However, from the consistency between the experimental data and the fit treating $M_{II}$ across $T_{C,2}$ as a phase transition, we believe that the "effective" symmetry breaking field is rather weak and does not distort the temperature dependence away from the order-parameter like behavior.

In summary, there are contributions from both $M_{I}$ and $M_{II}$ to the overall intensity (i.e., mainly $\chi_{M_I}^{EQ\,(i)}$ and $\chi_{M_{II}}^{EQ\,(i)}$) and the RA orientation (i.e., mainly $\chi_{M_I}^{EQ\,(c)}$ and $\chi_{M_{II}}^{EQ\,(c)}$). Using the Co kagome magnetic lattice to perform the symmetry analysis, we can see subsequent broken symmetries across $T_{C,1}$ and $T_{C,2}$ for $M_{I}$ and $M_{II}$, respectively. Using the whole crystalline lattice for the symmetry analysis, we see broken symmetries across $T_{C,1}$ but not $T_{C,2}$, which leads to a strict statement that $M_{I}$ undergoes a phase transition across $T_{C,1}$ but $M_{II}$ experiences a cross-over across $T_{C,2}$. However, we would like to highlight that the "effective" symmetry breaking field on $M_{II}$ across $T_{C,2}$ is so weak that the temperature dependence is close enough to the order-parameter-like behavior.

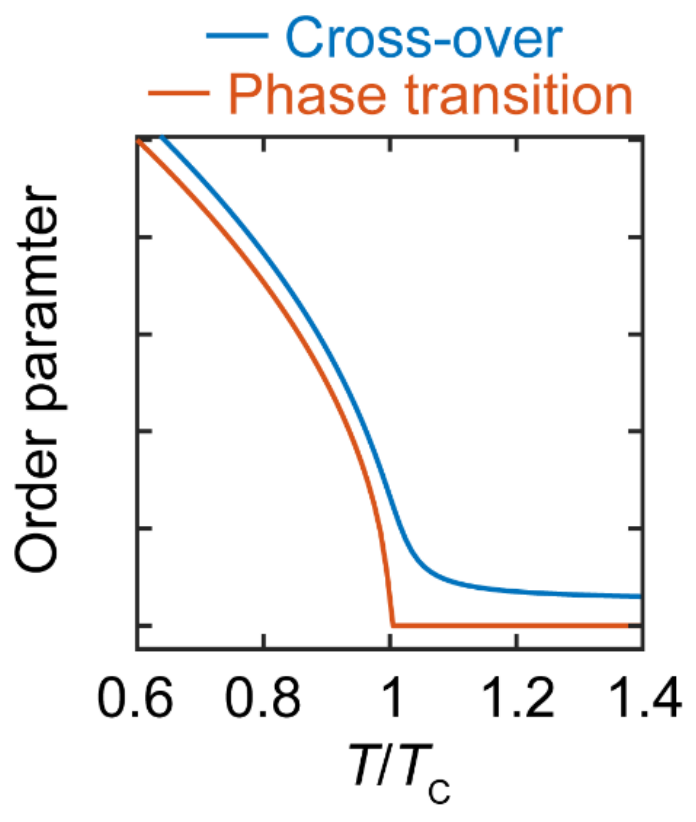

**Fig. S11** Order parameter as a function of normalized temperature $T/T_C$ for a second order phase transition and a cross-over behavior. The order parameter for the second order phase transition shows a continuous but unsmooth onset across the onset temperature of $T_C$. On the contrary, the cross-over behavior due to the presence of an "effective" field shows a continuous and smooth increase across $T_C$.